\RequirePackage{fix-cm}
\documentclass[11pt, a4]{article} 
\usepackage[ margin=1.0in]{geometry}
\usepackage{graphicx}
\usepackage[utf8]{inputenc}
\usepackage{fancyhdr}
\usepackage{xspace}
\usepackage{tabularx}
\usepackage{algorithm}
\usepackage{algorithmicx}
\usepackage{algcompatible}
\usepackage{algpseudocode}
\algrenewcommand\algorithmicindent{0.8em}%
\usepackage[toc,page]{appendix}
\usepackage{color}
\usepackage[font=scriptsize]{caption}
\usepackage{ragged2e}
\usepackage{tabularx}
\usepackage{amsmath,amssymb,latexsym} 
\usepackage{fancyvrb}
\usepackage{float}
\usepackage{cite}
\usepackage{epstopdf}
\usepackage{epsfig}
\usepackage{graphicx}
\usepackage{pgfplots}
\usepackage{lipsum}
\usepackage{booktabs,chemformula}
\usepackage{tabularx,booktabs} 
\usepackage{etoolbox}
\usepackage{breqn}
\usepackage{subfiles}
\usepackage{sidecap}
\usepackage{authblk}
\usepackage{soul}
\usepackage{mathtools}
\usepackage{pdfpages}
\usepackage{xspace}
\pgfplotsset{compat=1.11,
        /pgfplots/ybar legend/.style={
        /pgfplots/legend image code/.code={%
        \draw[##1,/tikz/.cd,bar width=3pt,yshift=-0.2em,bar shift=0pt]
                plot coordinates {(0cm,0.8em)};},
},
}
\usepackage{calligra}
\usepackage{calrsfs}
\DeclareMathAlphabet{\mathcalligra}{T1}{calligra}{m}{n}
\DeclareCaptionFont{tiny}{\tiny}
\algrenewcommand\alglinenumber[1]{\tiny #1:}
\usepackage{multicol}
\usepackage{comment}
\usepackage{subcaption}
\usepackage{graphicx}
\usepackage{authblk}
\usepackage{appendix}
\usepackage{xspace}
\usepackage{enumitem}
\usepackage{soul}
\usepackage{mathtools}
\usepackage{breqn}
\usepackage[hidelinks]{hyperref}
\usepackage{url}
\usepackage{listings}
\usepackage{abstract}
\usepackage{todonotes}
 
\newcommand{\punt}[1]{}
\newcommand{\cmnt}[1]{}

\newcommand{\cgds} {concurrent graph data-structure\xspace}

\newcommand{\lble} {linearizable\xspace}
\newcommand{\lbty} {linearizability\xspace}
\newcommand{\rbty} {reachability\xspace}

\newtheorem{theorem}{Theorem}
\newtheorem{lemma}{Lemma}

\newcounter{history}

\newenvironment{proof}[1][Proof]{\noindent\textbf{#1.} }{} 
\newenvironment{proofsketch}[1][Proof Sketch]{\noindent#1: }{\hfill $\Box$\\[0.4mm]} 

\newcommand{\secref}[1]{Section~\ref{sec:#1}}
\newcommand{\figref}[1]{Figure~\ref{fig:#1}}

\newcommand{\thmref}[1]{Theorem~\ref{thm:#1}}
\newcommand{\lemref}[1]{Lemma~\ref{lem:#1}}

\newcommand{\lineref}[1]{Line~\ref{lin:#1}}

\newcommand{\ignore}[1]{}

\algdef{SE}[DOWHILE]{Do}{doWhile}{\algorithmicdo}[1]{\algorithmicwhile\ #1}
\newcommand{\op} {operation\xspace}
\newcommand{\mth} {operation\xspace}

\newcommand{\lp} {LP\xspace}

\newcommand{\tru} {\texttt{true}\xspace}
\newcommand{\fal} {\texttt{false}\xspace}
\newcommand{\nul} {\texttt{NULL}\xspace}

\newcommand{\vnodes} {{\tt VNodes}\xspace}
\newcommand{\enodes} {{\tt ENodes}\xspace}
\newcommand{\enode}{{\tt ENode}\xspace}
\newcommand{\vnode}{{\tt VNode}\xspace}
\newcommand{\bfsnode}{{\tt BFSNode}\xspace}

\newcommand{\bfstree}{{\tt BFS\text{-}tree}\xspace}

\newcommand{\bfstrees}{{\tt BFS\text{-}trees}\xspace}
\newcommand{\vlist} {vertex-list\xspace}
\newcommand{\elist} {edge-list\xspace}
\newcommand{\elists} {edge-lists\xspace}

\newcommand{\vh}{\texttt{vh}\xspace}
\newcommand{\vt}{\texttt{vt}\xspace}
\newcommand{\eh}{\texttt{eh}\xspace}
\newcommand{\et}{\texttt{et}\xspace}

\newcommand{\fadd}{FetchAndAdd\xspace}

\newcommand{\Rcbl}{\textsc{Reachable}\xspace}

\newcommand{\rcbl}{{reachable}\xspace}
\newcommand{\scrcbl}{\textsc{SCRcbl}}
\newcommand{\dcrcbl}{\textsc{DCRcbl}}
\newcommand{\getpath}{\textsc{GetPath}\xspace}

\newcommand{\visitedarray}{\texttt{VisitedArray}\xspace}

\newcommand{\bfstclt}{\textsc{BFSTreeCollect}\xspace}

\newcommand{\of}{obstruction-free\xspace}
\newcommand{\Of}{Obstruction-free\xspace}

\newcommand{\nbk}{non-blocking\xspace}

\newcommand{\cas}{compare-and-swap\xspace}
\newcommand{\CAS}{\texttt{CAS}\xspace}
\newcommand{\tas}{test-and-set\xspace}
\newcommand{\faa}{fetch-and-add\xspace}
\newcommand{\FAA}{\texttt{FAA}\xspace}

\newcommand{\convplus}{\textsc{AcyConVPlus}\xspace}
\newcommand{\createbfs} {\textsc{AcyCBnode}\xspace}
\newcommand{\createe} {\textsc{AcyCEnode}\xspace}
\newcommand{\createv}{\textsc{AcyCVnode}\xspace}

\newcommand{\locvplus}{\textsc{AcyLocV}\xspace}
\newcommand{\loceplus}{\textsc{AcyLocE}\xspace}
\newcommand{\loccplus}{\textsc{AcyLocC}\xspace}
\newcommand{\concplus}{\textsc{AcyConCPlus}\xspace}

\newcommand{\acaddv}{\textsc{AcyAddV\xspace}}
\newcommand{\acremv}{\textsc{AcyRemV\xspace}}
\newcommand{\acadde}{\textsc{AcyAddE\xspace}}
\newcommand{\acreme}{\textsc{AcyRemE\xspace}}
\newcommand{\acconv}{\textsc{AcyConV\xspace}}
\newcommand{\accone}{\textsc{AcyConE\xspace}}

\newcommand{\ds}{data-structure\xspace}

\newcommand{\lf}{lock-free\xspace}
\newcommand{\wf}{wait-free\xspace}
\newcommand{\Wf}{Wait-free\xspace}

\newcommand{\enext}{{\tt enext}\xspace}
\newcommand{\vnext}{{\tt vnext}\xspace}
\newcommand{\bnext}{{\tt next}\xspace}

\newcommand{\pointv}{{\tt pointv}\xspace}
\newcommand{\ecount}{{\tt ecount}\xspace}
\newcommand{\lecount}{{\tt lecount}\xspace}

\newcommand{\vntp}{{\tt VERTEX NOT PRESENT}\xspace}
\newcommand{\entp}{{\tt EDGE NOT PRESENT}\xspace}
\newcommand{\ventp}{{\tt VERTEX OR EDGE NOT PRESENT}}

\newcommand{\ep}{{\tt EDGE PRESENT}\xspace}
\newcommand{\eap}{{\tt EDGE ALREADY PRESENT}\xspace}
\newcommand{\eadd}{{\tt EDGE ADDED}\xspace}

\newcommand{\er}{{\tt EDGE REMOVED}\xspace}

\newcommand{\cycle}{{\tt CYCLE}\xspace}

\newcommand{\scan}{\textsc{Scan}\xspace}

\newcommand{\SColt}{\textsc{Single Collect}\xspace}
\newcommand{\DColt}{\textsc{Double Collect}\xspace}

\newcommand{\comparepath}{\textsc{ComparePath}\xspace}
\newcommand{\comparetree}{\textsc{CompareTree}\xspace}

\newcommand{\isTransit}{\textsc{isTransit}\xspace}
\newcommand{\isMarked}{\textsc{isMarked}\xspace}
\newcommand{\MarkedRef}{\textsc{MarkedRef}\xspace}
\newcommand{\unMarkedRef}{\textsc{UnMarkedRef}\xspace}
\newcommand{\addedRef}{\textsc{AddedRef}\xspace}
\newcommand{\tranRef}{\textsc{TransitRef}\xspace}
\newcommand{\added} {\texttt{ADDED}\xspace}
\newcommand{\marked} {\texttt{MARKED}\xspace}
\newcommand{\transit} {\texttt{TRANSIT}\xspace}

\newcommand{\ops} {operations\xspace}

\newcommand{\sgt} {SGT\xspace}

\definecolor{butter1}{rgb}{0.988,0.914,0.310}
\definecolor{chocolate1}{rgb}{0.914,0.725,0.431}
\definecolor{chameleon1}{rgb}{0.541,0.886,0.204}
\definecolor{skyblue1}{rgb}{0.447,0.624,0.812}
\definecolor{plum1}{rgb}{0.678,0.498,0.659}
\definecolor{scarletred1}{rgb}{0.937,0.161,0.161}

\begin{document}

\title{A Pragmatic Non-Blocking Concurrent Directed Acyclic Graph 
}


\author{
     Sathya Peri$^1$,Muktikanta Sa$^2$, Nandini Singhal$^3$ \\
      Department of Computer Science \& Engineering \\
      Indian Institute of Technology Hyderabad, India \\
      \{$^1$sathya\_p, $^2$cs15resch11012\}@iith.ac.in, $^3$nandini12396@gmail.com
}

\date{}

\maketitle

\begin{abstract}


In this paper, we have developed two algorithms for maintaining acyclicity in a concurrent directed graph. The first algorithm is based on a wait-free reachability query and the second one is based on partial snapshot-based obstruction-free reachability query. Interestingly, we are able to achieve the acyclic property in the dynamic setting without the need of helping using descriptors by other threads or clean double collect mechanism. We present a proof to show that the graph remains acyclic at all times in the concurrent setting. We also prove that the acyclic graph data-structure operations are linearizable. We implement both the algorithms in C++ and test through a number of micro-benchmarks. Our experimental results show an average of $7$x improvement over the sequential and global lock implementation.

\textbf{keywords: }{acyclic graph \and concurrent data structure \and linearizability \and lock-freedom.}
\end{abstract}

\section{Introduction}
\vspace{-0.1in}
A graph is a common \ds that can model many real-world objects and pairwise relationships among them. Graphs have a huge number of applications in various field like social networking, VLSI design, road network, graphics, blockchains and many more. Usually, these graphs are \textit{dynamic} in nature, that is, they undergo dynamic changes like addition and removal of vertices and/or edges \cite{Demetrescu+:DynGraph::book:2004}. These applications also need \ds which supports dynamic changes and can expand at run-time depending on the availability of memory in the machine. 

Nowadays, multi-core systems have become ubiquitous. To fully harness the computational power of these systems, it has become necessary to design efficient data-structures which can be executed by multiple threads concurrently. In the past decade, there have been several efforts to port sequential data-structures to a concurrent setting, like stacks, queues, sets, trees. 

Most of these \ds use locks to handle mutual exclusion while doing any concurrent modifications. However, in an asynchronous shared-memory system, where an arbitrary delay or a crash failure of a thread is possible, a lock-based implementation is vulnerable to arbitrary delays or deadlock. For instance, a thread could acquire a lock and then sleep (or swapped out) for a long time. Or the thread could be involved cyclic wait with other threads while obtaining locks or crash after obtaining the lock. 

On the other hand, in a \lf \ds, threads do not acquire locks. Instead, they use atomic hardware instructions such as \cas, \tas etc. These instructions ensure that at least one non-faulty thread is guaranteed to finish its operation in a finite number of steps. Therefore, lock-free data-structures are highly scalable and naturally fault-tolerant.


Although several concurrent data-structures have been developed, concurrent graph data-structures and the related \op{s} are still largely unexplored. In several graph applications, one of the crucial requirements is preserving \textit{acyclicity}. Acyclic graphs are often applied to problems related to databases, data processing, scheduling, finding the best route in navigation, data compression, blockchains etc. Applications relying on graphs mostly use a sequential implementation and the access to the shared data-structures are synchronized through the global locks, which causes serious performance bottlenecks. 

A relevant application is \emph{Serialization Graph Testing (\sgt)} in Databases \cite[Chap 4]{Weikum+:TIS:book:2002} and Transactional Memory (TM) \cite{Sinha+:STM:ipdps:2010}. \sgt requires maintaining an acyclic graph on all concurrently executing (database or TM) transactions with edges between the nodes representing conflicts among them. In a concurrent scenario, where multiple threads perform different operations, maintaining acyclicity without using locks is not a trivial task. Indeed, it requires every shared memory access to be checked for the violation of the acylic property, which necessitates optimization of the steps in all these operations.

Apart from \sgt, maintaining acyclic graphs can be very useful in blockchains. Several popular blockchains maintain acyclic graphs such as tree structure (Bitcoin \cite{Brito+:bitcoin:2013}, Ethereum \cite{Buterin:Ethereum:2013} etc.) or general DAGs (Tangle \cite{Popov:Tangle:2018})). 

\subsection{Contributions}
In this paper, we present an efficient \nbk concurrent acyclic directed graph \ds  and  its \ops are similar to Chatterjee et. al. \cite{Chatterjee+:NbGraph:ICDCN-19} with some elegant modifications. The contributions of our work are summarized below:
\begin{enumerate}
	\item We describe an Abstract Data Type (ADT) that maintains an acyclic directed graph $G = (V,E)$. It comprises of the following \mth{s} on the sets $V$ and $E$: (1) Add Vertex: $\acaddv$ (2) Remove Vertex: $\acremv$, (3) Contains Vertex: $\acconv$ (4) Add Edge: $\acadde$ (5) Remove Edge: $\acreme$ and (6) Contains Edge: $\accone$. The ADT remains acyclic after completion of any of the above operations in $G$. The acyclic graph is represented as an adjacency list like in \cite{Chatterjee+:NbGraph:ICDCN-19}. 
	\item We present an efficient concurrent \nbk implementation of the ADT (\secref{ds}). We present two approaches for maintaining acyclicity: the first one is based on a \wf \rbty query and the second one is based on \of \rbty query similar to the \getpath \mth of Chatterjee et al.\cite{Chatterjee+:NbGraph:ICDCN-19} (\secref{algorithms}).
	
	\item We prove the correctness by showing the operations of the concurrent acyclic graph data-structure are \lble \cite{HerlWing:TPLS:1990}. We also prove the  non-blocking progress guarantee: (a) The operations $\acconv$ and $\accone$ are \wf, only if the vertex keys are finite; (b) Among the two algorithms for maintaining acyclicity, we show that the first algorithm based on searchability is \wf, whereas the second algorithm based on \rbty queries is \of and (c) The operations $\acaddv$, $\acremv$, $\acconv$, $\acadde$, $\acreme$, and $\accone$ are \lf. \secref{proof}.
	\item  We evaluated the \nbk algorithms in C++ implementation and tested through a number of micro-benchmarks. Our experimental results depict on an average of $7$x improvement over the sequential and global lock implementation (\secref{results}).
\end{enumerate}
\vspace{-0.2in}

\subsection{Related Work}
Kallimanis and Kanellou \cite{Kallimanis+:WFGraph:opodis:2015} presented a concurrent graph that supports \wf edge updates and traversals. They represent the graph using adjacency matrix, with bounded number of vertices. As a result, their graph data-structure does not allow any insertion or deletion of vertices after initialization of the graph. This may not be adequate for many real-world applications which need dynamic modifications of vertices as well as unbounded graph size. 

A recent work by Chatterjee et al. \cite{Chatterjee+:NbGraph:ICDCN-19} proposed a non-blocking \cgds which allowed multiple threads to perform dynamic insertion and deletion of vertices/edges. Our paper extends this \ds to maintain acyclicity of a directed graph. Their algorithm does not allow the the graph to maintain acyclicity after any update operation. 


\subsection{Overview of the Algorithm Design}
\label{subsec:overview}
Before getting into the technical details (in \secref{ds}) of the algorithm, we first provide an overview of the design. We implement an acyclic concurrent unbounded directed graph based on both existing \lf linked-list\cite{Harris:NBList:disc:2001} and the \cgds\cite{Chatterjee+:NbGraph:ICDCN-19}. The \textit{vertex-nodes} are placed in a sorted linked-list and the neighboring vertices of each vertex-node are placed in a rooted sorted linked-list of \textit{edge-nodes}. To achieve efficient graph traversal, we maintain a pointer from each edge-node to its corresponding vertex-node. Each vertex-node's \elist and \vlist are \lf with concurrent updates and lookup operations. 

As we know that lock-freedom is not composable \cite{Dang+:progress:disc:2011} and our algorithm is a composition of \lf operations, we prove the liveness of our algorithm independent of \lf list arguments. In addition to that, we also propose some graceful optimizations for the concurrent acyclic graph operations that not only enhance the performance but also simplify the design.

Our main requirement is preserving \textit{acyclicity} and one can see that a cycle is created only after inserting an edge to the graph. So, after inserting a new edge to the graph we verify if the resulting graph is acyclic or not. If it creates a cycle, we simply delete that edge from the graph. However, the challenge is that these intermediate steps must be oblivious to the user and the graph must always appear to be acyclic. We ensure this by adding a \emph{transit} field to the edges that are temporarily added. 

To verify the acyclic property of the graph we propose two efficient algorithms: first one based on a \wf \rbty query and the second one based on \of \rbty query similar to the \getpath operation of \cite{Chatterjee+:NbGraph:ICDCN-19}. Both the \rbty algorithms perform the breadth-first search (BFS) traversal. For the sake of efficiency, we implement BFS traversal in a non-recursive manner. However, in order to achieve the overall performance we do not help the \rbty queries.

\label{sec:intro}
\section{System Model and Preliminaries}\label{sec:model}



\textbf{The Memory Model.} 
We consider an asynchronous shared-memory model with a finite set of $p$ processors accessed by a finite set of $n$ threads. The non-faulty threads communicate with each other by invoking methods on the shared objects. We execute our acyclic graph \ds on a shared-memory multi-core with multi-threading enabled which supports atomic \texttt{read}, \texttt{write}, \texttt{\faa} (\FAA)  and \texttt{\cas} (\CAS) instructions. 


A \FAA${(x, a)}$ instruction atomically increments the value at the memory location ${x}$ by the value $a$. Similarly, a \texttt{CAS}${(x, a, a')}$ is an atomic instruction that checks if the current value at a memory location ${x}$ is equivalent to the given value ${a}$, and only if true, changes the value of ${x}$ to the new value ${a'}$ and returns \tru; otherwise the memory location remains unchanged and the instruction returns \fal. Such a system can be perfectly realized by a Non-Uniform Memory Access (NUMA) computer with one or more multi-processor CPUs.

\noindent
\textbf{Correctness.} We consider \textit{\lbty} proposed by Herlihy \& Wing \cite{HerlWing:TPLS:1990} as the correctness criterion for the graph operations. We assume that the execution generated by a data-structure is a collection of \mth invocation and response events. Each invocation of a method call has a subsequent response. An execution is \lble if it is possible to assign an atomic event as a \emph{linearization point} (\emph{\lp}) inside the execution interval of each \mth such that the result of each of these \mth{s} is the same as it would be in a sequential execution in which the \mth{s} are ordered by their \lp{s} \cite{HerlWing:TPLS:1990}. 

\noindent
\textbf{Progress.} The \emph{progress} properties specifies when a thread invoking \mth{s} on the shared memory objects completes in presence of other concurrent threads. In this context, we provide an acyclic graph implementation with operations that satisfies \emph{lock-freedom}, based on the definitions in Herlihy \& Shavit \cite{Herlihy+:OnNatProg:opodis:2001}.
\section{The Data Structure}\label{sec:ds}
\subsection{Abstract Data Type}
An acyclic graph is defined as a directed graph $G = (V, E)$, where $V$ is the set of vertices and $E$ is the set of directed edges. Each edge in $E$ is an ordered pair of vertices belonging to $V$. A vertex $v$ $\in V$ has an immutable unique key $k$ denoted by $v(k)$. A directed edge from the vertex $v(k_1)$ to $v(k_2)$ is denoted as $e(v(k_1), v(k_2))$ $\in E$. For simplicity, we denote $e(v(k_1), v(k_2))$ as $e(k_2)$, which means the $v(k_1)$ has a neighbouring vertex $v(k_2)$. 
\ignore{
\begin{minipage}{\linewidth}
  \centering
\begin{minipage}{.466\textwidth}
An acyclic graph is defined as a directed graph $G = (V, E)$, where $V$ is the set of vertices and $E$ is the set of directed edges. Each edge in $E$ is an ordered pair of vertices belonging to $V$. A vertex $v$ $\in V$ has an immutable unique key $k$ denoted by $v(k)$. A directed edge from the vertex $v(k_1)$ to $v(k_2)$ is denoted as $e(v(k_1), v(k_2))$ $\in E$. For simplicity, we denote $e(v(k_1), v(k_2))$ as $e(k_2)$, which means the $v(k_1)$ has a neighbouring vertex $v(k_2)$. 


\end{minipage}
\hspace{0.05\linewidth}
\begin{minipage}{.466\textwidth}
\scriptsize
\begin{figure}[H]
\captionsetup{font=footnotesize}
	\centerline{\scalebox{0.45}{\input{figs/AcyclicGraph.pdf}}}
    \vspace{-2mm}
	\caption{(a) An acyclic graph (b) The concurrent acyclic graph representation of \ds for (a).}
    \label{fig:ac-Graph}
\end{figure}
\end{minipage}
\end{minipage}
}
\noindent For a concurrent acyclic graph, we define following ADT operations:
\begin{enumerate}
	\item The $\acaddv(k)$ adds a vertex $v(k)$ to $V$, only if $v(k) \notin V$ and then returns \tru, otherwise it returns \fal. 
	\item The $\acremv(k)$ deletes a vertex $v(k)$ from $V$, only if $v(k) \in V$ and then returns \tru, otherwise it returns \fal. Once a vertex $v(k)$ is deleted successfully all its outgoing and incoming edges also removed.
	\item The $\acconv(k)$ if $v(k) \in V$ it returns \tru, otherwise it returns \fal.
	
	\item The $\acadde(k_1, k_2)$ adds an edge $e(v(k_1), v(k_2))$ to $E$, only if $e(v(k_1)$, $v(k_2))$ $\notin$ $E$ and $v(k_1) \in V$ and $v(k_2) \in V$ and it does not create any cycle then it returns \eadd. If $v(k_1) \notin V$ or $v(k_2) \notin V$, it returns \vntp. If $e(v(k_1), v(k_2)) \in E$, it returns \eap.
	\item The $\acreme(k_1, k_2)$ deletes the edge $e(v(k_1)$, $v(k_2))$ from $E$, only if $e(v(k_1), v(k_2)) \in E$ and $v(k_1) \in V$ and $v(k_2) \in V$ then it returns \er. If $v(k_1) \notin V$ or $v(k_2) \notin V$, it returns \vntp. If $e(v(k_1), v(k_2)) \notin E$, it returns \entp.
	\item The $\accone(k_1,k_2)$ if $e(v(k_1), v(k_2)) \in E$ and $v(k_1) \in V$ and $v(k_2) \in V$ then it returns \ep, otherwise it returns \ventp.
\end{enumerate}
\vspace{-0.2in}

\begin{figure}[!t]
	\captionsetup{font=footnotesize}
	\begin{subfigure}{.2\textwidth}	
	\begin{footnotesize}
		\begin{tabbing}
			\hspace{0.1in} \= \hspace{0.1in} \= \hspace{0.1in} \=  \hspace{0.1in} \= \\
			\> {\bf class \vnode \{} \\
			\> \> \texttt{int}  $k$; 
			\\
			\> \> {\vnode~ \vnext;} 
			\\
			\> \> {\enode~ \enext;} 
			 \\
			 \> \> {int~ \ecount;} 
			 \\
			\> \> {int \visitedarray [];} 
			\\
			\> \} 
			\end{tabbing}
				\end{footnotesize}
			\end{subfigure}
			\begin{subfigure}{.2\textwidth}	
			\begin{footnotesize}
					\begin{tabbing}
						\hspace{0.1in} \= \hspace{0.1in} \= \hspace{0.1in} \=  \hspace{0.1in} \= \\
			\\
			\> {\bf class \enode \{} \\
			\> \> \texttt{int}  $l$;   
			\\
			\> \> {\vnode~ \pointv;} 
			\\
			\> \> {\enode~ \enext;} 
			\\
			\> \}\\
				\> {\bf class \bfsnode \{} \\
							\> \> {\tt \vnode~   $n$;} 
							\\
							\> \> {\tt \bfsnode~ \bnext, p;} 
							\\
							\> \> {int    \lecount;} 
			                \\
							
							\> \} \\
		\end{tabbing}
	\end{footnotesize}
	\end{subfigure}
		\begin{subfigure}{.45\textwidth}	
		\captionsetup{font=footnotesize}
	\centerline{\scalebox{0.35}{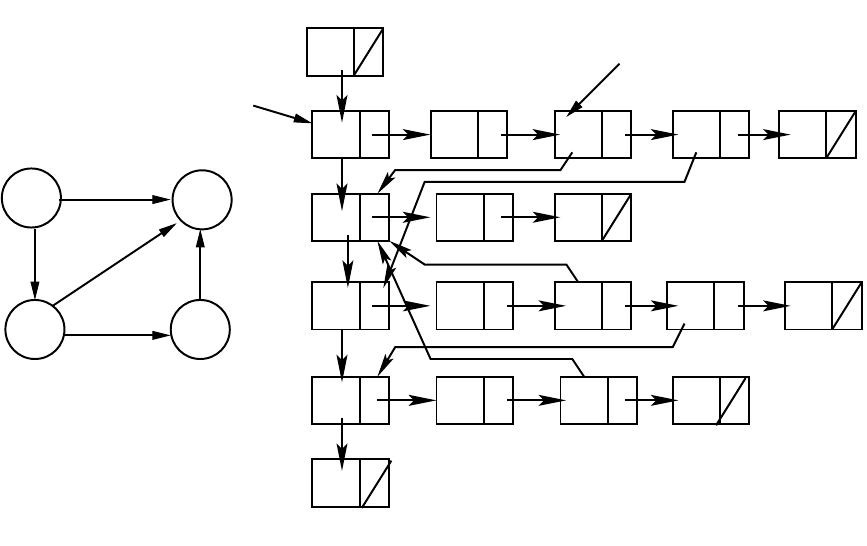}}
    \vspace{-2mm}
    \label{fig:ac-Graph}
			\ignore{
				\begin{footnotesize}
						\begin{tabbing}
					\hspace{0.1in} \= \hspace{0.1in} \= \hspace{0.1in} \=  \hspace{0.1in} \= \\
							\> {\bf class \bfsnode \{} \\
							\> \> {\tt \vnode~   $n$;} 
							\\
							\> \> {\tt \bfsnode~ \bnext;} 
							\\
							\> \> {\tt \bfsnode~ p;} 
							\\
							\> \> {int    \lecount;} 
			                \\
							
							\> \} \\
			\end{tabbing}
			\vspace{-0.3in}
		\end{footnotesize}
		}
		\end{subfigure}
		\vspace{-0.25in}
	   \setlength{\belowcaptionskip}{-15pt}
	\caption{Node structures used in the acyclic graph \ds: \enode, \vnode and \bfsnode. (a) An acyclic graph (b) The concurrent acyclic graph representation of \ds for (a).}
	\label{fig:ac-struct-evnode}
\end{figure}
\subsection{The \ds}
The proposed concurrent non-blocking acyclic graph \ds is constructed based on existing \lf linked-list\cite{Harris:NBList:disc:2001} and the \cgds\cite{Chatterjee+:NbGraph:ICDCN-19}. The \vnode, \enode and \bfsnode structures and the adjacency list representation of an acyclic graph are shown in \figref{ac-struct-evnode}. The \vnode structure has five fields, two pointers \vnext and \enext, an immutable key $k$, an atomic counter \ecount, and a \visitedarray array. The use of \ecount and \visitedarray is described in the later section. The pointer \vnext is an atomic pointer pointing to the next \vnode in the \vlist, whereas, an \enext pointer points to the edge head of the \elist of a \vnode. Similarly, an \enode structure has three fields, two pointers \enext and \pointv and an immutable key $l$. The \enext is an atomic pointer pointing to the next \enode in the \elist and the \pointv points to the corresponding \vnode, which helps direct access to its \vnode while doing any traversal like BFS, DFS, etc. We assume that all the \vnodes have a unique identification key $k$ and all the adjacency \enodes of a \vnode have also a unique key $l$. A \bfsnode has three pointers $n$, \bnext and $p$, and a counter \lecount. The pointer $n$ holds the corresponding \vnode's address, \bnext points to the next \bfsnode in the BFS-list and $p$ points to the corresponding parent. The local counter \lecount stores the $n$'s \ecount value which is used in the \comparetree and \comparepath methods.

We initialize the \vlist with dummy head(\vh) and tail(\vt) (called sentinels) with values $-\infty$ and $\infty$ respectively. Similarly, each \elists is also initialized with dummy head(\eh) and tail(\et)(see \figref{ac-struct-evnode}).

Our acyclic graph \ds maintains some \textit{invariants}: (a) the \vlist is sorted based on the \vnode's key value $k$ and each unmarked \vnode is reachable from the \vh, (b) also each of the \elists are sorted based on the \enode's key value $l$ and unmarked \enodes are reachable from the \eh of the corresponding \vnode and (c) the concurrent graph always stays \textit{acyclic}.

\section{Working of Non-blocking Algorithm}\label{sec:algorithms}
In this section, we describe the technical details of all acyclic graph operations. 


\noindent \textbf{Pseudo-code convention:} The acyclic graph algorithm is depicted in the \figref{ac-v-methods}, \ref{fig:ac-e-methods}, \ref{fig:ac-rcbl1-methods}, \ref{fig:ac-rcbl2-methods} and \ref{fig:e2-methods}. We use $p.x$ to access the member field $x$ of a class object pointer $p$.
To return multiple variables from an operation we use $\langle x_1, x_2,\ldots,x_n \rangle$. To avoid the overhead of another field in the node structure, we use bit-manipulation: last two significant bits of a pointer $p$. We define six methods $\isMarked(p)$ and $\isTransit(p)$, return \tru if last two significant bits of pointer $p$ are set to $01$ and $10$, respectively, else, both returns \fal, $\MarkedRef(p)$, $\unMarkedRef(p)$, $\addedRef(p)$ and $\tranRef(p)$ sets last two significant bits of the pointer $p$ to $01$, $00$, $11$ and $10$, respectively. An invocation of $\createv(k)$ creates a new \vnode with key $k$. Similarly, an invocation of $\createe(k)$ creates a new \enode with key $k$ with \transit state. Whereas, an invocation of $\createbfs(k)$ creates a new \bfsnode with a $v(k)$. For a newly created \vnode the pointer fields are \nul. Similarly, a newly created \enode initialises its pointer fields to \nul as well. In case of a new \bfsnode, the pointer field $n$, \bnext and $p$ are initialized with $v(k)$, \nul and parent node, respectively. Each slot of a \visitedarray in each \vnode is initialized to $0$ and the counter \ecount is also initialized to $0$.

To support acyclic property we modify the operation descriptor used by \cite{Chatterjee+:NbGraph:ICDCN-19} with a pointer in a single memory-word with \textit{bit-masking}. In case of an x86\text{-}64 bit architecture, memory has a 64-bit boundary and the last three least significant bits are unused. So, our operator descriptor uses the last two significant bit of the pointer. If the last two bits are set to $01$ the pointer is \marked, if it is set to $10$ the pointer indicates \transit, if it set to $11$ the pointer indicates \added and if it is set to $00$ the pointer is unused and unmarked.
 
\subsection{Acyclic Vertex Operations}
\begin{figure*}[!t]
	\begin{subfigure}{.53\textwidth}	
\begin{algorithmic}[1]
\renewcommand{\algorithmicprocedure}{\textbf{Operation}}
\scriptsize
	\Procedure{ \acaddv($key$)}{}\label{acadvstart}
	\While{(1)}
	\State{$\langle predv, currv \rangle$ $\gets$ \locvplus($\vh$, $key$);}\label{lin:acaddv-locv}
	\If{($currv.k$ $=$ $key$)}\label{lin:acaddv-cvk} 
	\State return {\fal;} 
	\Else
	\State{$newv$ $\gets$ \createv($key$); }
	\State{$newv.\vnext$ $\gets$ $currv$;  }
	\If{(CAS($predv.\vnext$, $currv, newv$))} \label{lin:cas-acaddv}
	\State return {\tru;} 
	\EndIf
	\EndIf
	\EndWhile
	\EndProcedure\label{acadvend}
	\algstore{addv}
\end{algorithmic}	
	    \hrule
    \begin{algorithmic}[1]
\renewcommand{\algorithmicprocedure}{\textbf{Operation}}
	\algrestore{addv}
	\scriptsize
	\Procedure{ $\acremv$($key$)}{}\label{acremvstart}
	\While{(1)}
	\State{$\langle predv, currv \rangle$ $\gets$ \locvplus($\vh$, $key$);}\label{lin:acremv-locv}
	\If{($currv.k$ $\neq$ $key$)}\label{lin:acremv-cvk}
	\State return {\fal;}
	\EndIf
	\State{$cnext$ $\gets$ $currv.\vnext$;}
	\If{($\neg$ \isMarked($cnext$))}
	\If{(CAS($currv.\vnext$, $cnext$, \MarkedRef($cnext$)))} \label{lin:cas-remv-lgic} 
	\If{(CAS($predv.\vnext, currv, cnext$))} \label{lin:cas-remv-phy}
	\State{ $break$;} 
	\EndIf
	\EndIf
	\EndIf
	\EndWhile
	\State return {\tru;}
	\EndProcedure\label{acremvend}
	\algstore{remv}
\end{algorithmic}
   	\end{subfigure}
    \begin{subfigure}{.45\textwidth}
	\begin{algorithmic}[1]
\renewcommand{\algorithmicprocedure}{\textbf{Operation}}
	\algrestore{remv}
	\scriptsize
	\Procedure{ $\acconv$($key$)}{}\label{acconvstart}
	\State {$currv$ $\gets$ $\vh.\vnext$;}
	\While{($currv.k$ $<$ $key$)}\label{lin:acconv-while}
	\State{$currv$ $\gets$ \unMarkedRef($currv.\vnext$);}
	\EndWhile \label{lin:acconv-while-end}
	\If{($currv.k$ $=$ $key$ $\bigwedge$ $\neg$ \isMarked($currv$))} 
	\State return  {\tru;}\label{lin:acconv-check}
	\Else 
	\State{return \fal;}
	\EndIf
	\EndProcedure\label{acconvend}
	\algstore{conv}
\end{algorithmic}
\hrule
\begin{algorithmic}[1]
	\algrestore{conv}
	\scriptsize
\renewcommand{\algorithmicprocedure}{\textbf{Operation}}	
	\Procedure{ $\accone$($k_1, k_2$)}{}\label{acconestart}
    	\State {$\langle$ u, v, st $\rangle$ $\gets$ $\concplus$($k_1, k_2);$}  \label{lin:accone-concplus}
		 \If{(st = \fal)}
        \State {return ``\vntp''; }
        \EndIf
        \State {$curre$ $\gets$ $u.\enext$;}
		\While{($curre.l$ $<$ $k_2$)} \label{lin:accone-while}
		\State{$curre$ $\gets$ \unMarkedRef($curre.\enext$);}
	    \EndWhile \label{lin:accone-while-end}
	    \If{($curre.l$ $=$ $k_2$ $\bigwedge$ $\neg$ \isMarked($u$) $\bigwedge$ $\neg$ \isMarked($v$) $\bigwedge$ $\neg$ \isMarked($curre$) $\bigwedge$ $\neg$ \isTransit($curre$))} \label{lin:accone-check}
        \State {return ``\ep'' ;}
        \Else
        \State{return``\ventp'';}
        \EndIf
	    \EndProcedure\label{acconeend}
		\algstore{cone}
\end{algorithmic}
	\end{subfigure}
	\setlength{\belowcaptionskip}{-15pt}
	\caption{Pseudo-codes of \acaddv, \acremv, $\acconv$ and \accone}\label{fig:ac-v-methods}
\end{figure*}

The acyclic vertex operations $\acaddv$, $\acremv$ and $\acconv$ depicted in \figref{ac-v-methods} are similar to the vertex operations in \cite{Chatterjee+:NbGraph:ICDCN-19}. The $\acconv$ do not help other threads in the process of traversal from the vertex head \vh to the destination vertex. If the vertex set keys are finite, then the $\acconv$ operation is \wf. 

An \acaddv($key$) operation is invoked by passing the  $key$ to be inserted, in Lines \ref{acadvstart} to \ref{acadvend}. It first traverses the \vlist in a \lf manner starting from \vh using $\locvplus$ procedure(\lineref{acaddv-locv}) until it finds a vertex with its key greater than or equal to $key$. In the process of traversal, it physically deletes all logically deleted \vnodes using a \CAS for helping a pending incompleted $\acremv$ operation. Once it reaches the appropriate location, say $currv$ and its predecessor, say $predv$, checks if the $key$ is already present. If the $key$ is not present it attempts a \CAS to add the new \vnode, say $newv$ in between the $predv$ and $currv$(\lineref{cas-acaddv}). On an unsuccessful \CAS, it is retried. 	

Like an \acaddv, an \acremv($key$) operation is invoked by passing the $key$ to be deleted, in  Lines \ref{acremvstart} to \ref{acremvend}. It traverses the \vlist in a \lf manner starting from \vh using \locvplus procedure(\lineref{acremv-locv}) until it finds a vertex with its key greater than or equal to $key$. Similar to the $\acaddv$ during the traversal it physically deletes all logically removed \vnodes using a \CAS for helping a pending incompleted $\acremv$ operation. Once it reaches the appropriate location, say $currv$ and its predecessor, say $predv$, checks $key$ already present earlier or not. If key is present it attempts to remove $currv$ in two steps (like \cite{Harris:NBList:disc:2001}), (a) atomically marks the \vnext of $currv$ using a \CAS(\lineref{cas-remv-lgic}), and (b) atomically updates the \vnext of the $predv$ to point to the \vnext of $currv$ using a \CAS(\lineref{cas-remv-phy}). On any unsuccessful \CAS, it reattempted.

When a vertex is deleted from a graph also needs to delete all the incoming and outgoing edges of it. Once a \CAS at \lineref{cas-remv-lgic} is successful, the vertex is logically deleted from the \vlist and its outgoing edges are deleted atomically. Notice that, all the incoming edges are logically deleted from the corresponding \enodes of any \elists. This is because each \enode has a direct pointer \pointv to its vertex node and calls \isMarked to validate the deleted \vnode. Finally, these \enodes are physically deleted using \CAS using any helping edge operations as described later.

An \acconv($key$) operation, in Lines \ref{acconvstart} to \ref{acconvend}, first traverses the \vlist in a \wf manner skipping all logically marked \vnodes until it finds a vertex with its key greater than or equal to $key$. Once it reaches the appropriate \vnode, checks its key value equal to $key$ and it is unmarked, then it returns \tru otherwise returns \fal. We do not allow $\acconv$ for any helping in the process of traversal.  
\vspace{-0.2in}
\subsection{Acyclic Edge Operations}

\begin{figure*}[!t]
 	\begin{subfigure}{.51\textwidth}	
	\begin{algorithmic}[1]
	\algrestore{cone}
\renewcommand{\algorithmicprocedure}{\textbf{Operation}}	
	\scriptsize
		\Procedure{ $\acadde$($k_1$, $k_2$)}{} \label{acaddestart}
		\State {$\langle$ u, v, st $\rangle$ $\gets$ $\convplus$($k_1, k_2);$}\label{lin:acadde-convplus}
		 \If{(st = \fal)}
        \State {return ``\vntp'' ; }
        \EndIf
        \While{($1$)}
	 \If{(\isMarked($u$) $\bigvee$  \isMarked($v$))}	\label{lin:acadde-check-uv}
        \State return ``\vntp''; 
        \EndIf
		\State{$\langle prede, curre \rangle$ $\gets$ \loceplus ($u.\enext$, $k_2$);} \label{lin:acadde-loceplus}
		\If{($curre.l$ $=$ $k_2$)} \label{lin:acadde-cvk}
		\State return ``\eap''; 
		\EndIf \label{lin:adde-endval-rcbl}
        \State{$newe$ $\gets$ \createe($k_2$);} 
		\State{$newe.\enext$ $\gets$ \tranRef($curre$);} 
		\State{$newe.\pointv$ $\gets$ $v$;} 
		\State{$nnext$ $\gets$ $newe.\enext$;} 
		\If{(\CAS($prede.\enext$, $curre, newe $ ))}  \label{lin:acadde-cas-transit} 
		\If{($\neg \scrcbl$($v,u$))} // $\scrcbl$ or $\dcrcbl$ is invoked \label{lin:adde-rcbl}
	    \State{$newe.\enext$ $\gets$ $\addedRef(nnxt)$;}\label{lin:acadde-t-a} 
	    \State{$u.\ecount.\fadd(1)$;} // only if $\dcrcbl$ is invoked \label{lin:acadde-increm}
		\State return ``\eadd'';
		\Else
		\State{$newe.\enext$ $\gets$ $\MarkedRef(nnxt)$;}\label{lin:acadde-t-m} 
		\State return `` \cycle''
		\EndIf
		\EndIf
		\EndWhile
        \EndProcedure \label{acaddeend}
		\algstore{adde}
\end{algorithmic}
	\end{subfigure}
 \begin{subfigure}{.48\textwidth}
	\begin{algorithmic}[1]
	\algrestore{adde}
	\scriptsize
\renewcommand{\algorithmicprocedure}{\textbf{Operation}}	
		\Procedure{ $\acreme$($k_1$, $k_2$)}{}\label{acremestart}
    		\State {$\langle$ u, v, st $\rangle$ $\gets$ $\convplus$($k_1, k_2$);}  \label{lin:acreme-convplus}
		 \If {(st = \fal)}
	    \State {return ``\vntp''; }
       \EndIf
       	\While{($1$)}
       	        \If{(\isMarked($u$) $\bigvee$  \isMarked($v$))}\label{lin:acreme-check-uv}
        \State return ``\vntp''; 
        \EndIf
		\State{$\langle prede, curre \rangle$ $\gets$ \loceplus ($u.\enext$, $k_2$);} \label{lin:acreme-loceplus}
		\If{($curre.l$ $\neq$ $k_2$)}\label{lin:acreme-cel}
		\State return ``\entp''; 
		\EndIf
		\State{$cnt$ $\gets$ $curre.\enext$;}
		\If{($\neg$ \isMarked($cnt$))} 
		\If{(CAS($curre.\enext$, $cnt$, \MarkedRef($cnt$)))}\label{lin:cas-acreme-lgic}
		\State{$u.\ecount.\fadd(1)$;}// only if $\dcrcbl$ is invoked \label{lin:acreme-increm}
		 \If{(CAS($prede.\enext, curre, cnt$))}  { $break$;}\label{lin:cas-acreme-phy}
                \EndIf
		\EndIf
		\EndIf
		\EndWhile
		\State return ``\er''; 
	    \EndProcedure\label{acremeend}
        \algstore{reme}
\end{algorithmic}
	\end{subfigure}
	\setlength{\belowcaptionskip}{-15pt}
	\caption{Pseudo-codes of $\acadde$ and \acreme.}\label{fig:ac-e-methods}
\end{figure*}

The acyclic edge operations $\acadde$ and $\acreme$ are depicted in \figref{ac-e-methods} and $\accone$ is depicted in \figref{ac-v-methods}. They are similar to the edge operations in \cite{Chatterjee+:NbGraph:ICDCN-19} with more changes to maintain the acyclic property of the concurrent graph. 

An \acadde($k_1$, $k_2$) operation, in Lines \ref{acaddestart} to \ref{acaddeend}, begins by validating the presence of the $v(k_1)$ and $v(k_2)$ in the \vlist by invoking \convplus(\lineref{acadde-convplus}) and validating that both the vertices are unmarked(\lineref{acadde-check-uv}). If the validations fail, it returns \vntp. Once the validation succeeds, \loceplus is invoked(\lineref{acadde-loceplus}) to find the location to insert the $e(k_2)$ in the \elist of the $v(k_1)$. The \loceplus is similar to the helping method \locvplus, but in the traversal phase it physically deletes two kinds of logically deleted \enodes (to help a pending incompleted $\acadde$ or $\acreme$ operations): (a) the \enodes whose \vnode is logically deleted using a \CAS, and (b) the logically deleted \enodes using a \CAS. The \loceplus traverses the \elist until it finds an \enode with its key greater than or equal to $k_2$. Once it reaches the appropriate location, say $curre$ and its predecessor, say $prede$ checks key $k_2$ already present earlier or not. If the key is already present it simply returns \ep otherwise it attempts a \CAS to add the new $e(k_2)$ with \transit state in between the $prede$ and $curre$(\lineref{acadde-cas-transit}). On an unsuccessful \CAS, the operation is re-tried. 

Once the edge $e(k_2)$ is inserted successfully, it invokes a reachable method to test whether $e(k_2)$ creates a cycle or not. As mentioned earlier, we propose two algorithms to maintain the acyclic property. First one is the \wf reachable algorithm $\scrcbl$(Single Collect Reachability), and the second one is the \of reachable algorithm $\dcrcbl$(Double Collect Reachability). The detailed working of these algorithms are given in the subsequent sections.  If the edge $e(k_2)$ creates a cycle, we delete it by setting its state from \transit to \marked (\lineref{acadde-t-m}) and return \cycle. Otherwise, we set the state from \transit to \added(\lineref{acadde-t-a}) and return \eadd.

\begin{figure*}[!t]
    \begin{subfigure}{.46\textwidth}
	\begin{algorithmic}[1]
	\algrestore{reme}
\renewcommand{\algorithmicprocedure}{\textbf{Operation}}	
	\scriptsize
	\Procedure{\scrcbl($k_1$, $k_2$)}{}\label{acrcblstart}
        \State{tid $\gets$ this\_thread.get\_id();} 
	   \State{queue $<$\vnode$>$ Q;} 
         \State{cnt $\gets$ cnt $+$ 1; } 
         \State{u.visitedArray[tid] $\gets$ cnt ;} \label{lin:rcbl-visitedarray}
        \State{Qe.enque($k_1$);} 
        \While{($\neg$Q.empty())} 
        \State{\vnode  cvn $\gets$ Q.deque();} 
        \State{eh $\gets$ cvn.enext;} 
        \For{(\enode itn $\gets$ \eh.\enext to \et)}
       
        \If{($\neg$\isMarked(itn))}  
        \State{\vnode adjn $\gets$ itn.\pointv;} 
        \If{($\neg$\isMarked(adjn))} 
        \If{(adjn = $k_2$)}   
        \State{return \tru;} 
        \EndIf
        \If{($adjn[tid] \neq cnt$)}
        \State{adjn.\visitedarray[tid] $\gets$ cnt;}
        \State{Q.enque(adjn);} 
        \EndIf
        \EndIf
        \EndIf
        \EndFor
    \EndWhile
     \State{return \fal;}
		\EndProcedure\label{acrcblend}
		\algstore{getpath-ac}
\end{algorithmic}
		\end{subfigure}
 \begin{subfigure}{.51\textwidth}
	\begin{algorithmic}[1]
	\algrestore{getpath-ac}
\renewcommand{\algorithmicprocedure}{\textbf{Operation}}	
	\scriptsize
	\Procedure{\bfstclt($k_1$, $k_2$)}{}\label{acbfstcltstart}
	 \State{tid $\gets$ this\_thread.get\_id();} 
         \State{queue $<$\bfsnode$>$ Q;} 
         \State{cnt $\gets$ cnt $+$ 1; } 
         \State{u.visitedArray[tid] $\gets$ cnt ;} 
        \State{bNode $\gets$ \createbfs($k_1$, \nul, \nul, $k_1$.\ecount );}
        \State{bTree.Insert(bNode);}\label{lin:bfsTree:insert1} 
        \State{Q.enque(bNode);} 
        \While{($\neg$Q.empty())} 
        \State{\bfsnode  cvn $\gets$ Q.deque();} 
        \State{eh $\gets$ cvn.n.\enext;} 
        \For{(\enode itn $\gets$ \eh.\enext to \et)}
       
        \If{($\neg$\isMarked(itn))}  
        \State{\vnode adjn $\gets$ itn.\pointv;}
        \If{($\neg$\isMarked(adjn))} 
        \If{(adjn = $k_2$)}
        \State{bNode $\gets$ \createbfs(adjn, cvn, \nul, adjn.\ecount);}
         \State{bTree.Insert(bNode);} \label{lin:bfsTree:insert2}
        \State{return $\langle bTree, \tru \rangle$;} \label{lin:bfsTree-true}
        \EndIf
        \If{($adjn[tid] \neq cnt$)}
        \State{adjn.\visitedarray[tid] $\gets$ cnt;} 
        \State{bNode $\gets$ \createbfs(adjn, cvn, \nul, adjn.\ecount);} 
         \State{bTree.Insert(bNode);}\label{lin:bfsTree:insert3} 
        \State{Q.enque(bNode);} 
        
        \EndIf
        \EndIf
        \EndIf
        \EndFor
    \EndWhile
     \State{return $\langle bTree, \fal \rangle$;}\label{lin:bfsTree-false}
		\EndProcedure\label{acbfstcltend}
		\algstore{acbfstclt}
\end{algorithmic}
		\end{subfigure}
    \setlength{\belowcaptionskip}{-16pt}	
	\caption{Pseudo-codes of  $\scrcbl$ and \bfstclt.}\label{fig:ac-rcbl1-methods}
\end{figure*}

Like \acadde, an \acreme($k_1$,$k_2$) operation, in Lines \ref{acremestart} to \ref{acremeend}, first validates the presence of the corresponding \vnodes and unmarked. If the validations fail, it returns \vntp. Once the validation succeeds, it finds the location to delete the $e(k_2)$ in the \elist of the $v(k_1)$. Similar to the \acadde, in the traversal phase it also physically deletes two kinds of logically deleted \enodes: (a) the \enodes whose \vnode is logically deleted, and (b) the logically deleted \enodes. It traverses the \elist until it finds an \enode with its key greater than or equal to $k_2$.  Once it reaches the appropriate location, say $curre$ and its predecessor, say $prede$ checks key $k_2$ already present earlier or not.  If key is not present earlier it returns \entp otherwise it attempts to remove $curre$ in two steps, (a) atomically marks the \enext of $curre$ using a \CAS(\lineref{cas-acreme-lgic}), and (b) atomically updates the \enext of the $prede$ to point to the \enext of $curre$ using a \CAS(\lineref{cas-acreme-phy}). On any unsuccessful \CAS, it reattempted. After successful \CAS it returns \er.

Similarly, an \accone($k_1$,$k_2$) operation, in Lines \ref{acconestart} to \ref{acconeend}, validates the presence of the corresponding \vnodes. Then it  traverses the \elist of $v(k_1)$ in a \wf manner skipping all logically marked \enodes until it finds an edge with its key greater than or equal to $k_2$. Once it reaches the appropriate \enode, checks its key value equal to $k_2$ and it is unmarked and not in \transit state and also $v(k_1)$ and $v(k_2)$ are unmarked, then it returns \ep otherwise it returns \ventp. Like \acconv, we also do not allow $\accone$ for any helping thread in the process of traversal.

\subsection{\Wf \SColt \Rcbl Algorithm}
In this section, we describe one of our algorithms to detect the cycle of a concurrent graph in a \wf manner. As we mentioned earlier that a cycle can be formed only after adding an edge to the graph. The \scrcbl($k_1$,$k_2$) operation, in Lines \ref{acrcblstart} to \ref{acrcblend}, performs non-recursive BFS traversal starting from the vertex $v(k_1)$. A reader can refer \cite{Cormen+:IntroAlgo:2009} to know the working of the BFS traversals in graphs. In the process of BFS traversal, it explores the \vnodes which are reachable from $v(k_1)$ and unmarked. If it reaches $v(k_2)$, then it terminates by returning \tru to the $\acadde$ operation. Then $\acadde$ deletes $e(k_2)$ by setting \enext pointer from the $\transit$ state to \marked state and returns \cycle. If it is unable to reach $v(k_2)$ from $v(k_1)$ after exploring all reachable \vnodes through \transit or \added or unmarked \enodes, then it terminates by returning \fal to the $\acadde$ operation. Then $\acadde$ adds  $e(k_2)$ by setting \enext pointer from the $\transit$ state to \added state and then it returns \eadd, which preserves the acyclic property after \acadde.

\ignore{So, to support this acyclic property we modified the operation descriptor used by \cite{Chatterjee+:NbGraph:ICDCN-19} with a pointer in a single memory-word with \textit{bit-masking}. In case of a x86\text{-}64 bit architecture memory has 64-bit boundary and last three least significant bits are unused. So, our operator descriptor uses the last two significant bit of the pointer. If the last two bits is set to $01$ the pointer is \marked, if it is set to $10$ the pointer is \transit, if it set to $11$ the pointer is \added and if it is set to $00$ the pointer is clean and unmarked.
}
\ignore{
 \begin{figure}[!t]
\captionsetup{font=small}
\centerline{\scalebox{.6}{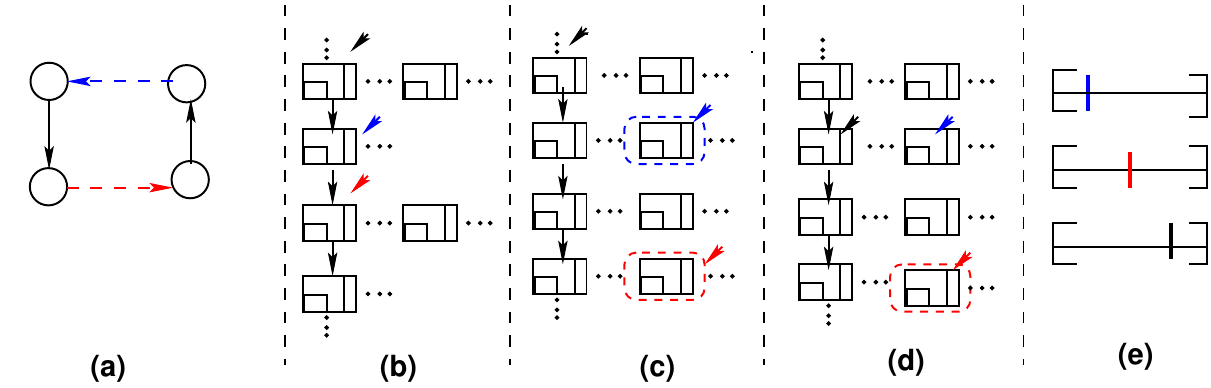}}
\setlength{\belowcaptionskip}{-15pt}
\caption{An example working of the methods while preserving acyclicity. (a) The initial graph, $T_1$, $T_2$ and $T_3$ are concurrently performing operations. The corresponding \ds is shown in (b). In (c), $T_3$ is traversing the vertex list, while $T_1$ and $T_2$ have added their corresponding edges in \transit, $T$ state and performing cycle detection. (d) $T_1$ has succeeded; and changed the status to \added, $A$. However, $T_2$ failed; it changes the status to \marked, $M$. Meanwhile, $T_3$ finds the respective edge. (e) One possible linearization of this concurrent execution.}
\label{fig:acyclic}
\end{figure}
}

In the process of BFS traversal, we used a \visitedarray( size the same as the number of threads) to put all the visited \vnodes locally. This is because multiple threads repeatedly invoke \rcbl operation concurrently, a boolean variable or a boolean array would not suffice like in case of sequential execution. We used a thread local variable $cnt$ as a counter for the number of repeated traversals by a thread. So, a \visitedarray slot keeps counter $cnt$ value (see \lineref{rcbl-visitedarray}).

However, an \enode in \transit state cannot be removed by any other concurrent thread other than the thread that added it, only if it creates a cycle. The threads which are performing cycle detection can see all the \enodes in \transit or \added state. Further, a concurrent $\accone$ operation will ignore all the \enodes with \transit state. This ensures that when an \enode is in the \added state an $\acadde$ operation return \eadd. \figref{acyclic} depicts an example of how acyclic property is preserved with concurrent $\acadde$ and $\accone$ execution and shows one possible \lp case.
\begin{figure*}[!t]
\captionsetup{font=footnotesize}
	\begin{subfigure}{.47\textwidth}	
\begin{algorithmic}[1]
	\algrestore{acbfstclt}
		\scriptsize
\renewcommand{\algorithmicprocedure}{\textbf{Operation}}	
		\Procedure{\dcrcbl($k_1$, $k_2$)}{}\label{acrcbl2start}
        \State{ tid $\gets$ this\_thread.get\_id();} 
         \State{ $status$ $\gets$ \scan($k_1$, $k_2$, $tid$);} \label{lin:rcbl-scan}
         \If{($status$ = \tru)}
          \State return {\tru;} 
         \Else 
         \State{return \nul; }
         \EndIf 
		\EndProcedure\label{acrcbl2end}
        \algstore{getpath}
\end{algorithmic}	
    \hrule
	\begin{algorithmic}[1]
	\algrestore{getpath}
		\scriptsize
		\Procedure{\scan($u$, $v$, $tid$)}{}\label{acscanstart}
         \State{list $<\bfsnode>$  otree, ntree ; } 
          \State{$\langle otree, of \rangle$ $\gets$ \bfstclt($u$, $v$, $tid$); }
        \While{(\tru)} \label{whilescan}
           \State{$\langle ntree, nf \rangle$ $\gets$ \bfstclt($u$, $v$, $tid$); } 
           \If{(of = \tru $\bigwedge$ nf = \tru $\bigwedge$ \comparepath(otree, ntree))}  \label{lin:scan-comparepath}
           \State{return $nf$;}
           \Else 
           \If{(of = \fal $\bigwedge$ nf = \fal $\bigwedge$ \comparetree(otree, nt))}\label{lin:scan-comparetree}
           \State{return nf;}
           \EndIf
           \EndIf
           \State{of $\gets$ nf;}
            \State{otree $\gets$ ntree;}
    \EndWhile
		\EndProcedure\label{acscanend}
        \algstore{scan}
\end{algorithmic}
	\end{subfigure}
 \begin{subfigure}{.51\textwidth}
	\begin{algorithmic}[1]
	\algrestore{scan}
		\scriptsize
		\Procedure{ \comparetree($otree, ntree$)}{}\label{ctreestart}
	 \If{(otree = \nul $\bigvee$ ntree = \nul)}
	     \State {return $\fal$ ; }
        \EndIf
        \State{\bfsnode oit $\gets$ otree.Head, nit $\gets$ ntree.Head;} 
        \Do 
        \If{(oit.n $\neq$ nit.n $\bigvee$ oit.\lecount $\neq$ nit.\lecount $\bigvee$ oldit.p $\neq$ newit.p)} {\hspace{2mm}return \fal; }
        \EndIf
        \State{oit $\gets$  oit.\bnext; nit $\gets$  nit.\bnext;}
        \doWhile{(oit $\neq$ ot.Tail $\bigwedge$ nit $\neq$ nt.Tail );} 
         \If{(oit.n $\neq$ nit.n $\bigvee$ oit.\lecount $\neq$ nit.\lecount $\bigvee$ oit.p $\neq$ nit.p)} {\hspace{2mm}return \fal; }
         \Else {\hspace{2mm}return $\tru$ ; }
         \EndIf
		\EndProcedure\label{ctreeend}
        \algstore{comparetree}
\end{algorithmic}
\hrule
\begin{algorithmic}[1]
	\algrestore{comparetree}
		\scriptsize
		\Procedure{ \comparepath($otree, ntree$)}{}\label{cpathstart}
	 \If{(otree = \nul $\bigvee$ ntree = \nul)}
	     \State {return $\fal$ ; }
        \EndIf
        \State{\bfsnode oit $\gets$ otree.Tail, nit $\gets$ ntree.Tail;}
        \Do 
         \If{(oit.n $\neq$ nit.n $\bigvee$ oit.\lecount $\neq$ nit.\lecount $\bigvee$ oldit.p $\neq$ newit.p)} {\hspace{2mm}return \fal; }
        \EndIf
        \State{oit $\gets$  oit.p; nit $\gets$  nit.p;}
        \doWhile{(oit $\neq$ ot.Head $\bigwedge$ nit $\neq$ nt.Head );} 
         \If{(oit.n $\neq$ nit.n $\bigvee$ oit.\lecount $\neq$ nit.\lecount $\bigvee$ oit.p $\neq$ nit.p)} {\hspace{2mm}return \fal; }
         
         \Else {\hspace{2mm}return \tru; }
         \EndIf
		\EndProcedure\label{cpathend}
        \algstore{comparepath}
	\end{algorithmic}
	\end{subfigure}
  \setlength{\belowcaptionskip}{-15pt}
	\caption{Pseudo-codes of \dcrcbl, \scan, $\comparetree$ and \comparepath.}\label{fig:ac-rcbl2-methods}
\end{figure*}

A side-effect to be observed here is that this may allow \textit{false positives}. This means that the algorithm may detect a cycle even though the graph does not contain one. This can happen in the following scenario; two threads $T_1$ and $T_2$ are adding edges lying in the path of a single cycle. In this case, both threads detect that the newly added \enode (in \transit state) has led to the formation of a cycle and both may delete their respective edges. However, in a sequential execution, only one of the edges will be removed. But, we allow this as the resulting graph at the end of each operation is still acyclic. In other words, false positives do not violate our correctness notion, \lbty. 

The proof of the acyclicity is given in   \lemref{addE-acyclic}(see \secref{app-correctness}). 

\subsection{\Of \DColt \Rcbl Algorithm}
To overcome the \textit{false positives} issues in case of $\scrcbl$ algorithm, we proposed an \of \rbty, $\dcrcbl$ algorithm, which is designed based on the atomic snapshot algorithm by Afek et al. \cite{Afek+:AtomicSnapshot:acmj:1993} and \rcbl algorithm by Chatterjee et al. \cite{Chatterjee+:NbGraph:ICDCN-19}. The \dcrcbl($k_1$,$k_2$) operation, in Lines \ref{acrcbl2start} to \ref{acrcbl2end}, performs \scan starting from $v(k_1)$. It checks whether $v(k_2)$ is reachable from $v(k_1)$. This reachable information is returned to the $\acadde$ operation and then $\acadde$ decides whether to add $e(k_2)$( is in the \transit state) to the \elist of $v(k_1)$ or not.

The \scan method, in Lines \ref{acscanstart} to \ref{acscanend}, first creates two \bfstrees, $otree$ and $ntree$ to hold the \vnodes in two consecutive BFS traversal. It performs repeated \bfstree collection by invoking \bfstclt until two consecutive collects are equal. The \bfstclt procedure, in Lines \ref{acbfstcltstart} to \ref{acbfstcltend}, performs non-recursive BFS traversal starting from the vertex $v(k_1)$. In the process of BFS traversal, it explores all reachable and unmarked \vnodes through adjacency \enodes which are in the \transit or \added or unmarked state. However, it keeps adding all these \vnodes in the $bTree$(see Line \ref{lin:bfsTree:insert1}, \ref{lin:bfsTree:insert2}, \ref{lin:bfsTree:insert3}). If it reaches $v(k_2)$, then it terminates by returning  $bTree$ and a reachable status \tru(\lineref{bfsTree-true}) to the \scan method. If it is unable to reach $v(k_2)$ from $v(k_1)$ after exploring all reachable \vnodes, then it terminates by returning $bTree$ and a reachable status \fal(\lineref{bfsTree-false}) to the \scan method.

If two consecutive \bfstclt method return the same boolean status value \tru then we invoke \comparepath to compare the two \bfstrees are equal or not. If both trees are equal, then \scan method returns \tru to $\dcrcbl$ operation, which says $v(k_2)$ is \rcbl from $v(k_1)$. Then $\dcrcbl$ returns \tru to the $\acadde$ operation and then $\acadde$ deletes that $e(k_2)$ by setting \enext pointer from the $\transit$  state to \marked state and returns \cycle, this is because $e(k_2)$ creates a cycle. However, if two consecutive \bfstclt methods return the same status value \fal then we invoke \comparetree to compare the two \bfstrees are equal or not. If it is, the \scan method returns \fal to the $\dcrcbl$ operation which says $v(k_2)$ is not \rcbl from $v(k_1)$. Then $\dcrcbl$ returns \fal to the $\acadde$ operation and then $\acadde$ adds that $e(k_2)$ by setting \enext pointer from the $\transit$ state to \added state and then it returns \eadd. Which confines the acyclic property after \acadde. If two consecutive \bfstclt method return the same boolean status value \tru or \fal but do not match in the \comparepath or \comparetree, then we discard the older \bfstree and restart the \bfstclt.

The \comparepath method, in Lines \ref{cpathstart} to \ref{cpathend}, compares two \bfstree based on path and along with the \lecount values. It starts from the last \bfsnode and follows the parent pointer $p$ until it reaches to the starting \bfsnode or any mismatch occurred at any \bfsnode. It returns \fal for any mismatch occurred otherwise returns \tru. Similarly, the \comparetree method, in Lines \ref{ctreestart} to \ref{ctreeend}, compares two \bfstree based on all explored \vnodes in the process of BFS traversal and along with the \lecount values. It starts from the starting \bfsnode and follows with the next pointer \bnext until it reaches the last \bfsnode or any mismatch occurred at any \bfsnode. It returns \fal for any mismatch occurred and otherwise returns \tru. We explain the necessary increment of an \ecount and requirement of a \lecount below. 

To capture the modifications along the path of BFS-traversal, we have an atomic counter \ecount associated with each vertex. During any edge update operations, before an $e(k_2)$ is physically deleted, the  counter \ecount of the source vertex $v(k_1)$ is certainly incremented at \lineref{acadde-increm} or \ref{lin:acreme-increm} or \ref{lin:faa:loc} either by the operation that logically deleted the $e(k_2)$ or any edge helping operations. To verify the double collect we compare with \bfstree alone with counter.

\section{Correctness and Progress Guarantee}\label{sec:proof}

In this section, we prove the correctness of our concurrent acyclic graph \ds based on  $\lp$\cite{HerlWing:TPLS:1990} events inside the execution interval of each of the operations.
\begin{theorem}\normalfont \label{thm:linzble}
The non-blocking concurrent acyclic graph operations are linearizable.
\end{theorem}

\begin{proof}
Based on the return values of the operations we discuss the \lp{s}.
\begin{enumerate}
	\item $\acaddv(k)$: Based on the return values we have two cases:
	\begin{enumerate}
		\item \tru: The \lp be the successful \CAS execution at the \lineref{cas-acaddv}.
		\item \fal: The \lp be the atomic read of the \vnext pointing to the vertex $v(k)$. 
	\end{enumerate}
	\item $\acremv(k)$: Based on the return values we have two cases:
	\begin{enumerate}
		\item \tru: The \lp be the successful \CAS execution at the \lineref{cas-remv-lgic} (logical removal).
		\item \fal: If there is a concurrent \acremv operation $op$, that removed $v(k)$ then the \lp be just after the \lp of $op$. If $v(k)$ did not exist in the \vlist then the \lp be at the invocation of $\acremv$.\label{step:acremv-fal} 
	\end{enumerate}
	\item $\acconv(k)$: Based on the return values we have two cases:
	\begin{enumerate}
		\item \tru: The \lp be the atomic read of the \vnext pointing to the vertex $v(k)$.
		\item \fal: The \lp be the same as returning \fal \acremv, the case \ref{step:acremv-fal}.
	\end{enumerate}

	\item $\acadde(k_1, k_2)$: Based on the return values we have four cases:
	\begin{enumerate}
		\item \eadd: \label{step:aceadd}
		\begin{enumerate}
			\item With no concurrent $\acremv(k_1)$ or $\acremv(k_2)$: The \lp be the successful \CAS execution at the \lineref{acadde-cas-transit}.
			\item With concurrent $\acremv(k_1)$ or $\acremv(k_2)$: The \lp be just before the first remove's \lp. 
		\end{enumerate}
		\item \cycle: 
		\begin{enumerate}
			\item With no concurrent $\acremv(k_1)$ or $\acremv(k_2)$: The \lp be the at the \lineref{acadde-t-m}.
			\item With concurrent $\acremv(k_1)$ or $\acremv(k_2)$: The \lp be just before the first remove's \lp. 
		\end{enumerate}
		\item \eap: 
		\begin{enumerate}
			\item With no concurrent $\acremv(k_1)$ or $\acremv(k_2)$: The \lp be the atomic read of the \enext pointing to the \enode $e(k_2)$ in the \elist fo the vertex $v(k_1)$. 
			\item With concurrent $\acremv(k_1)$ or $\acremv(k_2)$ or $\acreme(k_1, k_2)$ : The \lp be just before the first remove's \lp. 
		\end{enumerate}		
		\item \vntp: \label{step:acadde-vntp}
		\begin{enumerate}
			\item At the time of invocation of $\acadde(k_1, k_2)$ if both vertices $v(k_1)$ and $v(k_2)$ were in the \vlist and a concurrent \acremv removes $v(k_1)$ or $v(k_2)$ or both then the \lp be the just after the \lp of the earlier \acremv.
			\item At the time of invocation of $\acadde(k_1, k_2)$ if both vertices $v(k_1)$ and $v(k_2)$ were not present in the \vlist, then the \lp be the invocation point itself.
		\end{enumerate}
	\end{enumerate}	
					
	\item $\acreme(k_1, k_2)$:  Based on the return values we have three cases:
	\begin{enumerate}
		\item \er: 
		\begin{enumerate}
			\item With no concurrent $\acremv(k_1)$ or $\acremv(k_2)$: The \lp be the successful \CAS execution at the \lineref{cas-acreme-lgic}(logical removal).
			\item With concurrent $\acremv(k_1)$ or $\acremv(k_2)$: The \lp be just before the first remove's \lp. 
		\end{enumerate}
		\item \entp: If there is a concurrent \acreme operation removed $e(k_1,k_2)$ then the \lp be the just after its \lp, otherwise at the invocation of $\acreme(k_1, k_2)$ itself. 	
		\item \vntp: The \lp be the same as the case \acadde returning ``\vntp''\ref{step:acadde-vntp}.	
	\end{enumerate}			
	
	\item $\accone(k_1, k_2)$: Similar to \acreme, based on the return values we have three cases:
	\begin{enumerate}
		\item \ep:
		\begin{enumerate}
			\item With no concurrent $\acremv(k_1)$ or $\acremv(k_2)$: The \lp be the atomic read of the \enext pointing to the \enode $e(k_2)$ in the \elist fo the vertex $v(k_1)$. 
			\item With concurrent $\acremv(k_1)$ or $\acremv(k_2)$ or $\acreme(k_1, k_2)$ : The \lp be just before the first remove's \lp.
		\end{enumerate}
		\item \vntp: The \lp be the same as that of the \acadde's returning ``\vntp'' case.	
		\item \ventp: The \lp be the same as that of the \acreme's returning ``\entp'' and \acadde's returning ``\vntp'' cases.
	\end{enumerate}

\end{enumerate}
\noindent
From the above discussion once can notice that each operation's \lp{s} lies in the interval between the invocation and the return steps. For any invocation of an $\acaddv(k)$ operation the traversal terminates at the \vnode whose key is just less than or equal to $k$. Similar reasoning also true for invocation of an $\acadde(k_1, k_2)$ operation. Both the operations do the traversal in the sorted \vlist and \elist to make sure that a new \vnode or \enode does not break the invariant of the acyclic graph \ds. The \acremv and \acreme does not break the sorted order of the lists. Similarly the non update operations do not modify the \ds. Thus we concluded that all acyclic graph operations maintain the invariant across the \lp{s}.

This completes the proof of the \lbty.
\end{proof}

\vspace{-0.1in}
\begin{theorem}\normalfont \label{thm:correctness}
	The presented concurrent acyclic graph algorithm
	\begin{enumerate}[label=(\roman*)]
		\item The operations $\acconv$, $\accone$ and $\scrcbl$ are \wf, if the vertex keys is finite.\label{lflem1}
		\item The operation $\dcrcbl$ is \of. \label{lflem2}
		\item The operations $\acaddv$, $\acremv$, $\acconv$, $\acadde$, $\acreme$, and $\accone$ are \lf. \label{lflem3}
	\end{enumerate}
\end{theorem}

\begin{proof}
If the set of keys is finite, then the size of the acyclic graph has a fixed upper bound. This means there are only finite number of \vnodes in between sentinel \vh  and \vt in the \vlist. If any non-faulty thread invokes an \acconv, it terminates on reaching \vt with finite number of steps. Similar argument can be brought for an \accone operation. Also one can see that \scrcbl operation never returns true with concurrent update operations and it terminates with finite number of steps. This shows the \ref{lflem1}.

Like \scrcbl operation, a \dcrcbl, a \comparetree or a  \comparepath operations never returns true with concurrent update operations, which impose the \textbf{While} loop Line \ref{whilescan} in \scan method to not terminate. So, unless a non-faulty thread has taken the steps in isolation a \dcrcbl operation will never returns.  This shows the \ref{lflem2}. 

Whenever an insertion or a deletion operation is blocked by a concurrent delete operation by the way of a marked pointer, then that blocked operations is helped  to make safe return. Generally, insertion and lookup operations do not need help by a concurrent operation. So, if any random concurrent execution consist of any concurrent \ds operation, then at least one operation finishes its execution in a finite number of steps taken be a non-faulty thread. Therefore,  the acyclic graph operations \acaddv, \acremv, \acconv, \acadde, \acreme, and \accone are \lf. This shows the \ref{lflem3}.

\end{proof}

\section{Experimental Evaluation}
\vspace{-0.1in}
We performed our tests on 56 cores machine with Intel Xeon (R) CPU E5-2630 v4 running at 2.20 GHz frequency. Each core supports 2 logical threads. Every core's L1 has 64k, L2 has 256k cache memory are private to that core; L3 cache (25MB) is shared across all cores of a processor. The tests were performed in a controlled environment, where we were the sole users of the system. The implementation\footnote{The source code is available on https://github.com/PDCRL/ConcurrentGraphDS.} has been done in C++ (without any garbage collection) and threading is achieved by using Posix threads. All the programs were optimized at O3 level. 

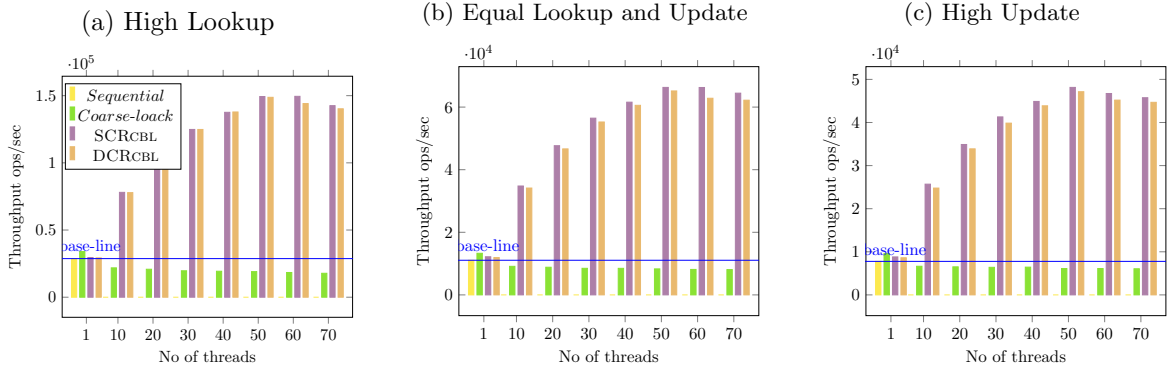
\begin{figure*}[tb]
\captionsetup{font=small}
\begin{subfigure}[b]{0.32\textwidth}
    \captionsetup{font=small, justification=centering}
   \caption{High Lookup}
    \centering
        \resizebox{\linewidth}{!}{
   	\begin{tikzpicture}[scale=0.230]
	\begin{axis}[
    ybar,
    xmin=1,
    xmax=70,
    enlargelimits=0.1,
   xticklabels={1, 10, 20, 30, 40, 50, 60, 70},
   xlabel near ticks,
   xtick=data,
    bar width=3.5pt,
    legend style={at={(0.21,0.97)},anchor=north},
	xlabel=No of threads,
	ylabel=Throughput ops/sec, 
    ylabel near ticks,]
	\addplot [butter1, fill=butter1] table [x=Threads, y=$Sequential$]{results/ds9010.dat};
		\addlegendentry{$Sequential$}
	\addplot[chameleon1,fill=chameleon1] table [x=Threads, y=$Coarse$]{results/ds9010.dat};
    	    \addlegendentry{$Coarse\text{-}loack$}
	\addplot [plum1,fill=plum1] table [x=Threads, y=$Single-Collect$]{results/ds9010.dat};
	 \addlegendentry{$\scrcbl$}
	\addplot [chocolate1,fill=chocolate1] table [x=Threads, y=$Double-Collect$]{results/ds9010.dat};
    \addlegendentry{$\dcrcbl$}
    \addplot[blue,sharp plot,update limits=false] 
	coordinates {(-15,28726) (90,28726)} 
	node[above] at (axis cs:2,29856) {base-line};
	\end{axis}
	\end{tikzpicture}
        }
        \label{fig:cgds-90-10}
    \end{subfigure}
    \begin{subfigure}[b]{0.32\textwidth}
    \setlength{\belowcaptionskip}{2mm}
    \captionsetup{font=footnotesize, justification=centering}
     \caption{Equal Lookup and Update}
    \centering
    \resizebox{\linewidth}{!}{
	\begin{tikzpicture} [scale=0.230]
	\begin{axis}[
    ybar,
    xmin=1,
    xmax=70,
    enlargelimits=0.1,
   xticklabels={1, 10, 20, 30, 40, 50, 60, 70},
   xlabel near ticks,
   xtick=data,
    bar width=3.5pt,
    legend style={at={(0.5,121,0.97)},anchor=north},
	xlabel=No of threads,
	ylabel=Throughput ops/sec,
    ylabel near ticks]
	\addplot [butter1, fill=butter1] table [x=Threads, y=$Sequential$]{results/ds5050.dat};
	\addplot[chameleon1,fill=chameleon1] table [x=Threads, y=$Coarse$]{results/ds5050.dat};
	\addplot[plum1,fill=plum1] table [x=Threads, y=$Single-Collect$]{results/ds5050.dat};
	\addplot[chocolate1,fill=chocolate1] table [x=Threads, y=$Double-Collect$]{results/ds5050.dat};
    \addplot[blue,sharp plot,update limits=false] 
	coordinates {(-15,11067) (90,11067)} 
	node[above] at (axis cs:2,12067) {base-line};
	\end{axis}
	\end{tikzpicture}
        }
        \label{fig:cgds-50-50}
    \end{subfigure}
    \begin{subfigure}[b]{0.32\textwidth}
    \captionsetup{font=footnotesize, justification=centering}
   \caption{High Update}
        \centering
        \resizebox{\linewidth}{!}{
           \begin{tikzpicture}[scale=0.230]
	\begin{axis}[
    ybar,
    xmin=1,
    xmax=70,
    enlargelimits=0.1,
   xticklabels={1, 10, 20, 30, 40, 50, 60, 70},
   xlabel near ticks,
   xtick=data,
    bar width=3.5pt,
    legend style={at={(0.621,0.6597)},anchor=north},
	xlabel=No of threads,
	ylabel=Throughput ops/sec,
    ylabel near ticks]
	\addplot [butter1, fill=butter1] table [x=Threads, y=$Sequential$]{results/ds1090.dat};
	\addplot[chameleon1,fill=chameleon1] table [x=Threads, y=$Coarse$]{results/ds1090.dat};
	\addplot[plum1,fill=plum1] table [x=Threads, y=$Single-Collect$]{results/ds1090.dat};
	\addplot[chocolate1,fill=chocolate1] table [x=Threads, y=$Double-Collect$]{results/ds1090.dat};
    \addplot[blue,sharp plot,update limits=false] 
	coordinates {(-15,7777) (90,7777)} 
	node[above] at (axis cs:2,7977) {base-line};
	\end{axis}
	\end{tikzpicture}
        }
        \label{fig:cgds-10-90}
    \end{subfigure}
    
\vspace{-6mm}
\setlength{\belowcaptionskip}{-15pt}
\caption{Acyclic Graph Data-Structure.} 
\label{fig:accgds}
\end{figure*}

We start our experiments by setting an initial directed graph with 1000 vertices and nearly $\binom{1000}{2}/4 \approx 125000$ edges added randomly. Then we create fixed number of threads with each thread randomly performing a set of operations chosen by a particular workload distribution. We evaluate the number of operations finished their execution in unit time and then calculate the throughput. We run each experiment for 20 seconds and the final data point values are collected after taking an average of 7 iterations. We present the results for the following workload distributions for acyclic directed graph over the ordered set of operations $\{$\acaddv, \acremv, \acconv, \acadde, \acreme, \accone$\}$ as: (1) \textit{High Lookup}:  $($$2.5\%$, $2.5\%$, $45\%$, $2.5\%$, $2.5\%$, $45\%$$)$, see the \figref{cgds-90-10}. (2) \textit{Equal Lookup and Update}: $($$12.5\%$, $12.5\%$, $25\%$, $12.5\%$, $12.5\%$, $25\%$$)$, see the \figref{cgds-50-50}. (3) \textit{High Update}: $($$22.5\%$, $22.5\%$, $5\%$, $22.5\%$, $22.5\%$, $5\%$$)$, \figref{cgds-10-90}. 

From \figref{accgds}, we notice that both $\scrcbl$ and $\dcrcbl$ algorithms perform well with the number of threads in comparison with sequential and coarse-lock based version. The \wf single collect reachable algorithm performs better than \of double collect reachable algorithm. However, the performance of the coarse-grained lock-based decreases with the number of threads. Moreover also it performs worse than even the sequential implementation. On average, both the non-blocking algorithms perform nearly $7\times$ times higher throughput over the sequential implementation. 


\label{sec:results}

\section{Conclusion}\label{sec:conc}
In this paper, we presented two efficient non-blocking concurrent algorithms for maintaining acyclicity in a directed graph where vertices and edges are dynamically inserted and/or deleted. The first algorithm is based on a wait-free reachability query and the second one is based on partial snapshot-based obstruction-free reachability query. Both these algorithms maintain the acyclic property of the graph throughout the concurrent execution. We prove that the acyclic graph data-structure operations are linearizable. We also present a proof to show that the graph remains acyclic at all times in the concurrent setting. We evaluated both the algorithms in C++ implementation and tested through a number of micro-benchmarks. Our experimental results performed on an average of $7$x improvement over the sequential implementation while the coarse lock implementation performs worse than the sequential one.

\bibliographystyle{plain}
\bibliography{biblio}

\begin{thebibliography}{10}

\bibitem{Afek+:AtomicSnapshot:acmj:1993}
Yehuda Afek, Hagit Attiya, Danny Dolev, Eli Gafni, Michael Merritt, and Nir
  Shavit.
\newblock {Atomic Snapshots of Shared Memory}.
\newblock {\em J. {ACM}}, 40(4):873--890, 1993.

\bibitem{Brito+:bitcoin:2013}
Jerry Brito and Andrea O'Sullivan.
\newblock {Bitcoin: A primer for policymakers}. mercatus center at george mason
  university,, 2013.

\bibitem{Buterin:Ethereum:2013}
V~Buterin.
\newblock {{Ethereum: a next generation smart contract and decentralized
  application platform}}, https://github.com/ethereum/wiki/wiki/white-paper,
  2013.

\bibitem{Chatterjee+:NbGraph:ICDCN-19}
Bapi Chatterjee, Sathya Peri, Muktikanta Sa, and Nandini Singhal.
\newblock {A Simple and Practical Concurrent Non-blocking Unbounded Graph with
  Linearizable Reachability Queries}.
\newblock In {\em Proceedings of the 20th International Conference on
  Distributed Computing and Networking, {ICDCN} 2019, Bangalore, India, January
  04-07, 2019}, pages 168--177, 2019.

\bibitem{Cormen+:IntroAlgo:2009}
Thomas~H Cormen, Charles~E Leiserson, Ronald~L Rivest, and Clifford Stein.
\newblock {\em {Introduction to algorithms}}.
\newblock MIT press, 2009.

\bibitem{Dang+:progress:disc:2011}
Nhan~Nguyen Dang and Philippas Tsigas.
\newblock {Progress Guarantees when Composing Lock-free Objects}.
\newblock In {\em Proceedings of the 17th International Conference on Parallel
  Processing - Volume Part II}, Euro-Par'11, pages 148--159, 2011.

\bibitem{Demetrescu+:DynGraph::book:2004}
Camil Demetrescu, Irene Finocchi, and Giuseppe~F. Italiano.
\newblock Dynamic graphs.
\newblock In {\em Handbook of Data Structures and Applications.} 2004.

\bibitem{Harris:NBList:disc:2001}
Timothy~L. Harris.
\newblock {A Pragmatic Implementation of Non-blocking Linked-Lists}.
\newblock In {\em Distributed Computing, 15th International Conference, {DISC}
  2001, Lisbon, Portugal, October 3-5, 2001, Proceedings}, pages 300--314,
  2001.

\bibitem{Herlihy+:OnNatProg:opodis:2001}
Maurice Herlihy and Nir Shavit.
\newblock {On the Nature of Progress}.
\newblock In {\em Principles of Distributed Systems - 15th International
  Conference, {OPODIS} 2011, Toulouse, France, December 13-16, 2011.
  Proceedings}, pages 313--328, 2011.

\bibitem{HerlWing:TPLS:1990}
Maurice~P. Herlihy and Jeannette~M. Wing.
\newblock Linearizability: a correctness condition for concurrent objects.
\newblock {\em ACM Trans. Program. Lang. Syst.}, 12(3):463--492, 1990.

\bibitem{Kallimanis+:WFGraph:opodis:2015}
Nikolaos~D. Kallimanis and Eleni Kanellou.
\newblock {Wait-Free Concurrent Graph Objects with Dynamic Traversals}.
\newblock In {\em 19th International Conference on Principles of Distributed
  Systems, {OPODIS} 2015, December 14-17, 2015, Rennes, France}, pages
  27:1--27:17, 2015.

\bibitem{Popov:Tangle:2018}
Serguei Popov.
\newblock The tangle, https://iota.org/iota whitepaper.pdf, 2018.

\bibitem{Sinha+:STM:ipdps:2010}
Arnab Sinha and Sharad Malik.
\newblock {Runtime checking of serializability in software transactional
  memory}.
\newblock In {\em 24th {IEEE} International Symposium on Parallel and
  Distributed Processing, {IPDPS} 2010, Atlanta, Georgia, USA, 19-23 April 2010
  - Conference Proceedings}, pages 1--12, 2010.

\bibitem{Weikum+:TIS:book:2002}
Gerhard Weikum and Gottfried Vossen.
\newblock {\em Transactional Information Systems: Theory, Algorithms, and the
  Practice of Concurrency Control and Recovery}.
\newblock Morgan Kaufmann, 2002.

\end{thebibliography}

\newpage
\appendix

\begin{center}
\Large\textbf{Appendix}
\end{center}
\label{sec:appendix-accg}
\ignore{
\section{Proof of the Correctness}
\label{sec:app-correctness}
\begin{theorem}\normalfont The non-blocking concurrent acyclic graph operations are linearizable.
\end{theorem}

\begin{proof}
Based on the return values of the operations we discuss the \lp{s}.

\noindent
From the above discussion once can notice that each operation's \lp{s} lies in the interval between the invocation and the return steps. For any invocation of an $\acaddv(k)$ operation the traversal terminates at the \vnode whose key is just less than or equal to $k$. Similar reasoning also true for invocation of an $\acadde(k_1, k_2)$ operation. Both the operations do the traversal in the sorted \vlist and \elist to make sure that a new \vnode or \enode does not break the invariant of the acyclic graph \ds. The \acremv and \acreme does not break the sorted order of the lists. Similarly, the non-update operations do not modify the \ds. Thus we concluded that all acyclic graph operations maintain the invariant across the \lp{s}.

This completes the proof of the \lbty.

\end{proof}

\begin{theorem}\normalfont
	The presented concurrent acyclic graph algorithm
	\begin{enumerate}[label=(\roman*)]
		\item The operations $\acconv$, $\accone$ and $\scrcbl$ are \wf, if the vertex keys is finite.\label{lflem1_a}
		\item The operation $\dcrcbl$ is \of. \label{lflem2_a}
		\item The operations $\acaddv$, $\acremv$, $\acconv$, $\acadde$, $\acreme$, and $\accone$ are \lf. \label{lflem3_a}
	\end{enumerate}
\end{theorem}
\begin{proof}
If the set of keys is finite, then the size of the acyclic graph has a fixed upper bound. This means there is an only finite number of \vnodes in between sentinel \vh  and \vt in the \vlist. If any non-faulty thread invokes an \acconv, it terminates on reaching \vt with a finite number of steps. A similar argument can be brought for an \accone operation. Also one can see that \scrcbl operation never returns true with concurrent update operations and it terminates with a finite number of steps. This shows the \ref{lflem1_a}.

Like \scrcbl operation, a \dcrcbl, a \comparetree or a  \comparepath operation never returns true with concurrent update operations, which impose the \textbf{While} loop Line \ref{whilescan} in \scan method to not terminate. So, unless a non-faulty thread has taken the steps in isolation a \dcrcbl operation will never return.  This shows the \ref{lflem2_a}. 

Whenever an insertion or a deletion operation is blocked by a concurrent delete operation by the way of a marked pointer, then that blocked operations is helped  to make a safe return. Generally, insertion and lookup operations do not need help by a concurrent operation. So, if any random concurrent execution consists of any concurrent \ds operation, then at least one operation finishes its execution in a finite number of steps taken be a non-faulty thread. Therefore,  the acyclic graph operations \acaddv, \acremv, \acconv, \acadde, \acreme, and \accone are \lf. This shows the \ref{lflem3_a}.

\end{proof}
}
\section{Proof of the Acyclicity}\label{sec:app-correctness}
\begin{lemma}
\label{lem:addE-acyclic}
In any state, all the edges with the status \added are acyclic. 
\end{lemma}

\begin{proofsketch}
Suppose $G$ is a graph which is acyclic and all its edges are in the \added state before invocation of  $\acadde(k,l)$ by any thread $T_x$. 

Let the $T_i$ and $T_j$ be the two threads concurrently invokes $\acadde(k_i,l_i)$ and $\acadde(k_j,l_j)$ respectively. Suppose  both the edges are in the same directed path $v(l)$ $\rightsquigarrow$ $v(k)$ and both independently create cycle with their respective edge. We use $e(k_i,l_i,st)$ to represent an edge, where $st$ is the current status of the edge, it may be \added, \transit or \marked.  

when $T_i$ invokes an $\acadde(k_i,l_i)$ operation, first it checks the validations from the  \lineref{acadde-convplus}-\ref{lin:adde-endval-rcbl}. After successfully validations $T_i$ adds the edge $e(k_i,l_i,\transit)$ and then invokes the $\scrcbl(l_i, k_i)$ to test cycle. Similarly, $T_j$ also added an $e(k_j,l_j, \transit)$ and invokes $\scrcbl(l_j, k_j)$ to check a cycle. Both the threads return boolean status and the \bfstree. The \bfstree contains the edges which are not \marked (means only the edges which are either \added or \transit). Now, we have three cases:

\begin{itemize}
\item $e(k_i,l_i, \transit)$ $\prec$  $e(k_j,l_j,\transit)$: Here $e(k_i,l_i, \transit)$ got added before $e(k_j,l_j, \transit)$. From our assumption that all the edges in the path $v(l)$ $\rightsquigarrow$ $v(k)$ are in the state \added and the graph $G$ is acyclic. Now $T_i$ added the edge  $e(k_i,l_i,\transit/\added)$ before  $T_j$ adds the edge $e(k_j,l_j,\transit)$ and then $T_j$ in the process of \scrcbl (in the \lineref{adde-rcbl}). The $T_j$ collects the \bfstree starting from  $v(l_j)$ and it will consider the edge $e(k_i,l_i,\transit/\added)$. If it reaches $v(k_j)$ by using the path $v(l_j)$ $\rightsquigarrow$ $v(k_j)$ it simply discards its edge by setting the state from \transit to \marked in \lineref{acadde-t-m}. Otherwise, it adds the edge by setting the state from \transit to \added in \lineref{acadde-t-a}.  Hence this case cycle is not possible and the $G$ remain as acyclic.

\item $e(k_j,l_j, \transit)$ $\prec$  $e(k_i,l_i,\transit)$: Here $e(k_j,l_j, \transit)$ got added before $e(k_i,l_i,\transit)$. It is similar to the previous case, from our assumption that all the edges in the path $v(l)$ $\rightsquigarrow$ $v(k)$ are in the state \added, all the vertices are reachable from the $v(l)$ and the graph $G$ is acyclic. Now $T_j$ added the edge  $e(k_j,l_j,\transit/\added)$ before  $T_i$ adds the edge $e(k_i,l_i,\transit)$ and then $T_i$ in the process of \scrcbl (in \lineref{adde-rcbl}) it will consider the edge  $e(k_j,l_j,\transit/\added)$. If it finds a path $v(l_i)$ $\rightsquigarrow$ $v(k_i)$ it simply discards its edge by setting the state from \transit to \marked in \lineref{acadde-t-m}. Hence this case cycle also not possible and the $G$ remain as acyclic.

\item $e(k_i,l_i, \transit)$ $\approx$  $e(k_j,l_j,\transit)$: Here $e(k_i,l_i, \transit)$ got added approximately same time $e(k_j,l_j,\transit)$. From our assumption that all the edges in the path $v(l)$ $\rightsquigarrow$ $v(k)$ are in the state \added, all the vertices are reachable from the $v(l)$ and the graph $G$ is acyclic. Now $T_i$ and $T_j$ added the edge  $e(k_i,l_i,\transit)$ and $e(k_j,l_j,\transit)$ at same time respectively. Then both $T_i$ and $T_j$ are in the process of \scrcbl (in \lineref{adde-rcbl}). $T_i$  will consider the edge  $e(k_j,l_j,\transit)$ added by $T_j$ and it finds there is a path $v(l_i)$ $\rightsquigarrow$ $v(k_i)$  and then discards its edge by setting the state from \transit to \marked in \lineref{acadde-t-m}. Similarly, $T_j$  will consider the edge  $e(k_i,l_i,\transit)$ added by $T_i$ and it finds there is a path $v(l_j)$ $\rightsquigarrow$ $v(k_j)$  and then discards its edge by setting the state from \transit to \marked in the \lineref{acadde-t-m}. Hence this case cycle also not possible and the $G$ remain as acyclic.
\end{itemize}
\end{proofsketch}
\section{The Acyclic Graph Algorithms }
 \begin{figure}[!t]
\captionsetup{font=small}
\centerline{\scalebox{.7}{\input{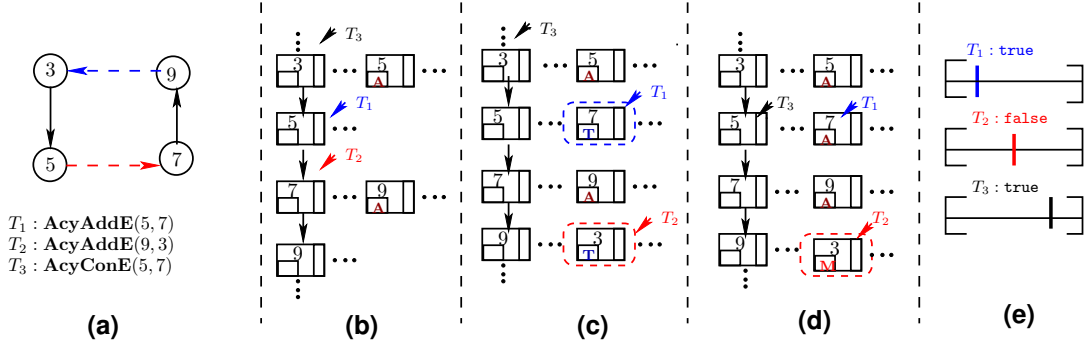}}}
\setlength{\belowcaptionskip}{-15pt}
\caption{An example working of the methods while preserving acyclicity. (a) The initial graph, $T_1$, $T_2$ and $T_3$ are concurrently performing operations. The corresponding \ds is shown in (b). In (c), $T_3$ is traversing the vertex list, while $T_1$ and $T_2$ have added their corresponding edges in \transit, $T$ state and performing cycle detection. (d) $T_1$ has succeeded; and changed the status to \added, $A$. However, $T_2$ failed; it changes the status to \marked, $M$. Meanwhile, $T_3$ finds the respective edge. (e) One possible linearization of this concurrent execution.}
\label{fig:acyclic}
\end{figure}
\label{sec:ac-app-code}
\begin{figure*}[!thb]
	\begin{subfigure}{.5\textwidth}	
\begin{algorithmic}[1]
	\algrestore{comparepath}
	\scriptsize
		\Procedure{$\locvplus$($v$, $k$)}{}\label{locvstart}
		\While{($\tru$)} \label{lin:search-again}
		\State {$predv \gets v$; $currv \gets predv.\vnext$;} \label{lin:locv3-w}
		\While{$(\tru)$}
		\State{$cn \gets currv.\vnext$;}
	    \While{($\isMarked(cn))$ $\bigwedge$ 
	    $(currv.k < k))$} \label{lin:locv5-w} 
		\If{($\neg$ CAS( $predv.\vnext$, $currv$, $currv.\vnext$))}
		\State goto \ref{lin:search-again}; 
		\EndIf
        \State {$currv \gets cn$;} 
        \State {$cn$ $\gets$ $\unMarkedRef(currv.\vnext)$;}
        \EndWhile
        \If{($currv.k \geq k$)} 
        \State{return $\langle predv, currv \rangle$}
        \EndIf
        \State {$predv \gets currv$; $currv \gets cn$;} 
		\EndWhile
        \EndWhile
		\EndProcedure\label{locvend}
		\algstore{locvplus}
\end{algorithmic}
	    \hrule
   \begin{algorithmic}[1]
	\algrestore{locvplus}
	\scriptsize
		\Procedure{$\loceplus$($v$, $k$)}{}\label{locestart}
		\While{(\tru)}  \label{lin:retry-locte}
		\State {$prede \gets v$; $curre \gets prede.\enext$;} 
        \While{(\tru)}
        \State{$cnt \gets curre.\enext$; \vnode  vn $\gets$ curre.\pointv;} 
        \While{(\isMarked($vn$) $\bigwedge$ $\neg$ \isMarked($cnt$))} \label{lin:retry-locte2}
		\If{($\neg$CAS($curre.\enext$, $cnt$, \MarkedRef($cnt$)))} \label{lin:loce-log-mark}
		\State{goto Line \ref{lin:retry-locte};}
		\EndIf
		\If{($\neg$CAS($prede.\enext, curre, cnt$))}\label{lin:loce-phy} {goto Line \ref{lin:retry-locte};}  
		\EndIf
        \State {$curre \gets cnt$; $ vn \gets curre.\pointv;$} 
        \State {$cnt \gets \unMarkedRef(curre.\enext)$;}
		\EndWhile
        \While{(\isMarked(cnt))}
		\State{v.\ecount.\fadd(1);}\label{lin:faa:loc}
		\If{($\neg$ CAS($prede.\enext, curre, cnt))$;}\label{remlogd} {goto \ref{lin:retry-locte}; }
		\EndIf
        \State {$curre \gets cnt$;  $ vn \gets curre.\pointv;$} 
        \State {$cnt \gets \unMarkedRef(curre.\enext)$;}
        \EndWhile
        \If{(\isMarked($vn$))} { goto Line \ref{lin:retry-locte2};}
        \EndIf
        \If{($curre.l \geq k$)} {return $\langle prede, curre \rangle$} 
        
        \EndIf
        \State {$prede \gets curre$; $curre \gets cnt$;} 
		\EndWhile
		\EndWhile
        \EndProcedure\label{loceend}
		\algstore{locte}
\end{algorithmic}
   	
	\end{subfigure}
    \begin{subfigure}{.5\textwidth}
\begin{algorithmic}[1]
	\algrestore{locte}
	\scriptsize
		\Procedure{$\convplus$ ($k_1$, $k_2$)}{}\label{convpstart}
		\If{($k_1 < k_2$)}
		\State {$\langle predv1, currv1\rangle$ $\gets$ $\locvplus$($\vh$, $k_1$);} \label{lin:locvplus-k}
		\If{($currv1.k$ $\neq$ $k_1$)}
	     \State{return $\langle \nul, \nul, \fal \rangle$;}
		\EndIf
		\State {$\langle predv2, currv2\rangle$ $\gets$ $\locvplus$($currv1$, $k_2$);} \label{lin:locvplus-l}
		\If{($currv2.k$ $\neq$ $k_2$)}
	     \State{return $\langle \nul, \nul, \fal \rangle$;}
		\EndIf
		\Else
		\State {$\langle predv2, currv2\rangle$ $\gets$ $\locvplus$($\vh$, $k_2$);}
		\If{($currv2.k$ $\neq$ $k_2$)}
	     \State{return $\langle \nul, \nul, \fal \rangle$;}
		\EndIf
		\State {$\langle predv1, currv1\rangle$ $\gets$ $\locvplus$($currv2$, $k_1$);}
		\If{($currv1.k$ $\neq$ $k_1$)}
	     \State{return $\langle \nul, \nul, \fal \rangle$ ;} 
		\EndIf
		\EndIf
		\State {returns $\langle currv1, currv2, \tru \rangle$;} 
        \EndProcedure\label{convpend}
		\algstore{convplus}
\end{algorithmic}	
\hrule
\begin{algorithmic}[1]
	\algrestore{convplus}
	\scriptsize
		\Procedure{$\loccplus$($v$, $k$)}{}
	\State {$predv \gets v$; $currv \gets predv.\vnext$;} \label{lin:locv3-start}
		\While{($\tru$)} \label{lin:loccplus-search-again}
		\If{($currv.k \geq k$)}
        \State{return $\langle predv, currv \rangle$;}
        \EndIf
        \State {$predv \gets currv$; $currv \gets \unMarkedRef(currv.\vnext)$;} 
		\EndWhile
		\EndProcedure
		\algstore{loccplus}
	\end{algorithmic}
	\hrule
	\begin{algorithmic}[1]
	\algrestore{loccplus}
	\scriptsize
		\Procedure{$\concplus$ ($k_1$, $k_2$)}{}
		\If{($k_1 < k_2$)}
		\State {$\langle predv1, currv1\rangle$ $\gets$ $\loccplus$($\vh$, $k_1$);}
		\If{($currv1.k$ $\neq$ $k_1$)}
	     \State{return $\langle \nul, \nul, \fal \rangle$;}
		\EndIf
		\State {$\langle predv2, currv2\rangle$ $\gets$ $\loccplus$($currv1$, $k_2$);}
		\If{($currv2.k$ $\neq$ $k_2$)}
	     \State{return $\langle \nul, \nul, \fal \rangle$;}
		\EndIf
		\Else
		\State {$\langle predv2, currv2\rangle$ $\gets$ $\loccplus$($\vh$, $k_2$);}
		\If{($currv2.k$ $\neq$ $k_2$)}
	     \State{return $\langle \nul, \nul, \fal \rangle$;}
		\EndIf
		\State {$\langle predv1, currv1\rangle$ $\gets$ $\loccplus$($currv2$, $k_1$);}
		\If{($currv1.k$ $\neq$ $k_1$)}
	     \State{return $\langle \nul, \nul, \fal \rangle$ ;} 
		\EndIf
		\EndIf
		\State {returns $\langle currv1, currv2, \tru \rangle$;} 
        \EndProcedure
\end{algorithmic}
	\end{subfigure}
\setlength{\belowcaptionskip}{-15pt}
	\caption{Pseudo-codes of \loceplus, \concplus, $\loccplus$ and \concplus.}\label{fig:e2-methods}
\end{figure*}

\end{document}